\documentclass[journal]{IEEEtran}
\IEEEoverridecommandlockouts
\usepackage{epsfig,amsmath,amssymb,amsfonts,bm,tikz,lipsum}
\usepackage{gensymb}
\usepackage{algorithm}
\usepackage{algorithmic}
\usepackage{cite}
\usepackage{CJK}
\usepackage{url}
\usepackage[percent]{overpic}
\usepackage{graphicx}
\usepackage{subfigure}
\usepackage{epstopdf}
\graphicspath{{figure/}}
\usepackage{amssymb}
\usepackage{textcomp}
\usepackage{xcolor}
\usepackage{cuted,flushend}
\usepackage[framemethod=tikz]{mdframed}
\usepackage{multirow}
\usepackage{booktabs}
\usepackage{framed}
\usepackage{float}
\usepackage{rotating}
\usepackage[breaklinks,colorlinks,linkcolor=black,citecolor=black,urlcolor=black]{hyperref}
\usepackage{flushend}
\usepackage{balance}
\usepackage{threeparttable}
\usepackage{makecell}
\allowdisplaybreaks[4]

\begin{document}
\title{Twofold Structured Features-Based Siamese Network for Infrared Target Tracking}
\author{Weijie~Yan, Guohua~Gu, Yunkai~Xu, Xiaofang~Kong, Ajun~Shao, Qian~Chen,  and Minjie~Wan
	\thanks{This work was supported in part by the National Natural Science Foundation of China  under Grant 62001234 and 62201260, in part by the Natural Science Foundation of Jiangsu Province  under Grant BK20200487, in part by the Fundamental Research Funds for the Central Universities under Grant 30923011015, and in part by the Equipment Pre-research Key Laboratory Fund Project under Grant 6142604210501. (Corresponding author: Minjie Wan.) }
   \thanks{Weijie Yan, Guohua Gu, Yunkai Xu, Ajun Shao, Qian Chen, and Minjie Wan are with the School of Electronic and Optical Engineering, Nanjing University of Science and Technology, Nanjing 210094, China, and also with the Jiangsu Key Laboratory of Spectral Imaging \& Intelligent Sense, Nanjing University of Science and Technology, Nanjing 210094, China. (email:  yanweijie@njust.edu.cn; gghnjust@mail.njust.edu.cn; xuyunkai@njust.edu.cn; njustsaj@njust.edu.cn; chenq@njust.edu.cn; minjiewan1992@njust.edu.cn.) }
   \thanks{Xiaofang Kong is with National Key Laboratory of Transient Physics, Nanjing University of Science and Technology, Nanjing 210094, China. (e-mail: xiaofangkong@njust.edu.cn).}
}

\maketitle
\begin{abstract}
Nowadays, infrared target tracking has been a critical technology in the field of computer vision and has many applications, such as urban security, pedestrian counting, smoke and fire detection, and so forth. Unfortunately, due to the absence of detailed information such as texture or color, it is easy for tracking drift to occur when the tracker encounters infrared targets that vary in shape or size. In order to address this issue, we present a twofold structured features-based Siamese network for infrared target tracking. Above all, a novel feature fusion network is proposed to make full use of both shallow spatial information and deep semantic information in a comprehensive manner, so as to improve the discriminative capacity for infrared targets. Then, a multi-template update module is designed to effectively deal with interferences from target appearance changes which are prone to cause early tracking failures. Finally, both qualitative and quantitative experiments are implemented on VOT-TIR 2016 and GTOT datasets, which demonstrates that our method achieves the balance of promising tracking performance and real-time tracking speed against other state-of-the-art trackers.
\end{abstract}


\IEEEpeerreviewmaketitle
\section{Introduction}
\IEEEPARstart{O}{wing} to the rapid development of artificial intelligence nowadays, infrared target tracking has become one of the frontier issues in the field of computer vision and shows promising potential for many computational social system-related tasks ~\cite{1-liu2022learning}~\cite{2-du2021overview}~\cite{wan2021infrared}, e.g., urban security ~\cite{3-abaya2014low}, pedestrian counting ~\cite{4-kristoffersen2016pedestrian}, smoke and fire detection ~\cite{5-gaur2020video}, and so forth ~\cite{6-cao2022tctrack}. Meanwhile, with the steady advancement of thermal imaging technology, infrared target tracking, which is capable of working under a variety of complicated illumination conditions, has been widely explored over the last few decades. In this paper, we are dedicated to designing a novel tracker for infrared target tracking that is flexibly adaptable to complex scenes.

Because of the promising accuracy and robustness, Siamese network-based trackers have been regarded as one of the most popular trackers in the field of visual object tracking ~\cite{zhang2022object}. Bertinetto et al. first ~\cite{7-bertinetto2016fully} introduced a fully-convolutional Siamese network trained end-to-end and solved the tracking problem by calculating the similarity between input features, which greatly improved the accuracy. Since then, numerous target tracking methods based on Siamese networks have emerged. For example, considering the limitations of shallow feature extraction network, Li et al. ~\cite{8-li2019siamrpn++} tried to apply a modern depth network called ResNet as the backbone network, which makes a further step in extracting specific and comprehensive target features. Generally speaking, the existing Siamese trackers employ modern depth network as their backbone and apply either fixed templates or simple linear variable templates to calculate the similarity with the search region.

Despite the advantages of these Siamese trackers in handling visible images ~\cite{zhang2023learning}, their tracking performance is rarely satisfactory when directly applied to treat infrared images. It is apparent that infrared images have their own particular limitations that lead to the degradation of tracking performance in Siamese trackers. First of all, due to the poor imaging quality of thermal detectors, infrared images are neither rich in texture, color and many other details, nor do they have high image resolution and signal-to-noise ratio, which makes it difficult for the tracker to distinguish the target from the background. To make it worse, target features extracted by modern depth network always have the tendency of focusing too much on deep semantic information and thereby ignoring shallow spatial information unconsciously ~\cite{9-he2018twofold}, which is equally critical to infrared target tracking. Inspired by these problems, we hope to design a more reasonable backbone to overcome the drawback of modern depth network and thus to better extract the features of the target from the background in the infrared image. Furthermore, infrared image is a kind of gray-scale image and is vulnerable to the surrounding environment~\cite{6-cao2022tctrack}, resulting in extensive changes in contour shape, gray-scale distribution, and other information of the target's appearance ~\cite{10-zhang2019learning}. Due to using fixed templates or simple linear variable templates, Siamese trackers are ill-equipped to resolve the apparent changes of the target that frequently arise in infrared images. Therefore, we aspire to develop a more effective method that allows the template features extracted by our Siamese tracker to adapt promptly as the infrared target changes.

In this paper, we propose a twofold structured features-based Siamese tracker equipped with the multi-template update module, named \textbf{TSF-SiamMU}, to overcome the two above-mentioned drawbacks of conventional depth network-based Siamese trackers when handling infrared images. First of all, a new feature fusion network, which separately fuses shallow spatial information and deep semantic information into the extracted features, is designed so as to greatly enhance the network’s capacity in distinguishing the fuzzy infrared target from the background. Further, we develop a template update mechanism which aims to estimate the current optimal template through the aggregation of the initial template, the accumulated template and the current template based on the previous prediction. Finally, in order to deal with various infrared target changes, a multi-template update module based on the template update mechanism is proposed, where differentiated update operations are applied to templates in diverse depth. By implementing both qualitative and quantitative experiments on real infrared sequences, the results indicate that our proposed approach achieves real-time performance and outperforms other state-of-the-art methods in terms of precision and success rate. In summary, this paper's principal contributions can be summarized below in three main aspects:

(1) A novel feature fusion network which differentiates the extracted features into shallow spatial features and deep semantic features is proposed to enhance the feature representation capacity of the infrared target in a comprehensive manner.

(2) A multi-template update module based on the template update mechanism is designed to further tackle the problem of tracking drift caused by the interference from various appearance changes of infrared targets.

(3) Experiments based on real infrared sequences prove that TSF-SiamMU not only enables advanced tracking results but also runs at 47 FPS on average, achieving the balance between precision and speed.

Four sections will be included in the following article. In Section~\ref{section:2}, a concise overview of existing target tracking methods will be provided. Section~\ref{section:3} will give a detailed explanation of our tracking model. In Section~\ref{section:4}, we aim to establish the credibility of our approach by conducting ablation and comparison experiments. Section~\ref{section:5} will summarize the proposed tracker.

\section{Related Work}\label{section:2}
In accordance with the approach to modelling target appearance, the current research on target tracking methods can be divided into two categories: trackers based on generative models and trackers based on discriminative models. Typical generative trackers such as Kalman filter ~\cite{11-zhang2010new}, particle filter ~\cite{12-chang2005kernel} and mean shift ~\cite{13-comaniciu2002mean}, apply online feature learning to establish appearance models and then search for the region with minimum reconstruction error in subsequent frames as target position by template matching. However, such kind of methods based on generative models are inferior to those based on discriminative models in accuracy, as a result of both underutilization of image information and ignorance of background information. Discriminative trackers typically transform the tracking process into a binary classification process where distinguish the target from the background. Nowadays, there have been two dominant methods so far: correlation filtering (CF) methods and deep learning methods. Here, we would like to focus on discussing these two kinds of trackers as follows. 

As for the CF approaches ~\cite{zhang2022scstcf}, Bolme et al. ~\cite{14-bolme2010visual} first introduced the concept of correlation filtering to the field of target tracking and constructed a new filter which correlates target and subsequent image based on two-dimensional Gaussian distribution response, thus ensuring a substantial increase in tracking accuracy and tracking speed. Immediately afterwards, Henriques et al. ~\cite{15-henriques2014high} proposed the utilization of kernel functions to map ridge regression in linear space to high-dimensional nonlinear space, as well as a novel cyclic sampling structure to further achieve high-dimensional nonlinear classification under dense sampling, both of which significantly improved the tracking performance. Following this, various target tracking methods based on the traditional CF framework have emerged. Danelljan et al. ~\cite{16-danelljan2014accurate} adopted a multi-feature fusion mechanism based on KCF to train the scale filter and position filter for target scale estimation and target localization respectively, which can better cope with the scale changes occurring in tracking process. In general, although such CF algorithms are remarkable in terms of tracking speed, it is rather difficult for them to maintain the satisfactory tracking accuracy.
 
Though researchers had realized the crucial importance of robust and accurate features for the construction of a target appearance model, there was no alternative before to replace such manually designed features that always have flaws. Fortunately, with the development of deep learning, neural networks have gradually gained significant advantages in feature extraction, which exactly fits the requirements of model design in target tracking tasks. Ma et al. ~\cite{17-ma2015hierarchical} introduced deep learning into CF trackers for the first time and replaced histogram of oriented gradient (HOG) features with hierarchical convolutional features (HCFs) in the trained VGG-19, which finally localized the target by weighted feature maps that fuses features of various depth. Inspired by the structure of HCF, Danelljan et al. ~\cite{18-danelljan2016beyond} applied the implicit interpolation to expand feature channels of different resolutions to the higher dimensional continuous spatial domain, and then the object of interest is successfully localized by using Hessian matrix. Despite the advantages of overly complex feature combinations, they may reduce the running speed and increase the risk of overfitting. In order to increase tracking speed without sacrificing tracking accuracy, Danelljan et al. ~\cite{19-danelljan2017eco} then proposed a factorization convolution method to simplify the feature extraction dimension and a Gaussian mixture model to merge similar samples; they also designed a model updating strategy against shading challenges based on C-COT. 

Nevertheless, CF trackers still have significant limitations when handling targets with complex appearance ~\cite{zhang2021visual}, thus a new tracking framework is needed to exploit the full strength of deep learning when extracting features. Tao et al. ~\cite{20-tao2016siamese} were the first to propose the Siamese network-based tracker, which learns a matching function by offline training and then uses the function to locate the target by calculating the score of template frames and sampled search frames. To make the most of target features, Bertinetto et al. ~\cite{7-bertinetto2016fully} went a step further by using a novel fully-convolutional Siamese network and a tracking approach that employs both offline training and online tuning, which has become the basis for future improvements of Siamese network. In order to enhance the generalization performance of the network, He et al. ~\cite{21-he2018twofold} divided the antecedent feature extraction network into two different networks, named semantic branch and appearance branch, which are separately trained to keep the heterogeneity of the two types of features. To overcome the interference of scale variation, Li et al. ~\cite{22-li2018high} brought region proposal network (RPN) into Siamese network based on SiamFC, improving tracking quality by separating localization process into classification branch and regression branch. Considering that the shallow backbone network in SiamRPN is difficult to extract specific and comprehensive target features, a modern depth network called ResNet was applied by Li et al. ~\cite{8-li2019siamrpn++}, in which the translational invariance was overcome, and the problem of asymmetry in two branches was dealt with by utilizing uniform sampling and multiple layers. Focusing on getting rid of the inconsistence existing in both both classification and regression branches, Zhang et al. ~\cite{zhang2022siamoa} applied the intersection over union to guide the branches and further proposed an offset-ware regression branch to make the prediction of bounding boxes more accurate.

Although accuracy has been improved by introducing RPN into Siamese trackers, such anchor-based trackers wasted considerable time on model training and coped poorly with large scale variations owing to the addition of hyper-parameters of anchor boxes. To deal with the problem, Chen et al. ~\cite{23-chen2020siamese} proposed an anchor-free Siamese tracker which views tracking problems as a parallel classification and regression problem and thus directly classified objects and regress their bounding boxes. For the purpose of solving the issue of poor robustness of the anchor-free method, Zhang et al. ~\cite{24-zhang2020ocean} proposed a feature alignment-based tracker using different sampling strategies in classification and regression branches. Focusing on making the features extracted from the Siamese network more discriminative, Xu et al. ~\cite{25-xu2021hierarchical} applied a new hierarchical backbone network to take advantage of target feature information at various levels of depth, a channel attention module to reinforce the specific key channels of the target in a selective manner, and an adaptive update mechanism to further deal with numerous problems arising from the interference of appearance similarity. Considering the anchor-free Siamese tracker produces a large number of low-quality bounding boxes and makes the background interference more apparent, Zhang et al. ~\cite{zhang2023siamese} proposed an improved head network to filter out low-quality bounding boxes and a recurrent criss-cross attention module to make target features more discriminative.

With the development of feature extraction networks, the significance of more distinctive target features throughout the entire tracking process has increased over time. Since many extracted features are useless, Yu et al. ~\cite{26-yu2020deformable} introduced various attentional mechanisms to the Siamese tracker, applying not only self-attention module to enrich the contextual information, but also mutual attention module to interact the information between template and search region before the correlation operation. Enlightened by the occurrence of transformer model ~\cite{27-vaswani2017attention}, Chen et al. ~\cite{28-chen2021transformer} proposed a new feature fusion Siamese tracker based on attention mechanism. By repeatedly using the double cross-attention feature enhancement module for feature fusion and finally fusing the features together with the additional cross-feature enhancement module, the tracker addressed the challenge that inter-correlation operations lose semantic information and easily lead to local optimum instead of global optimum. For the purpose of fully exploring the temporal contexts which were largely overlooked by most trackers, Wang et al. ~\cite{wang2021transformer} designed a unique Siamese network by separating the encoder and decoder into two parallel branches instead of the common use of transformer for feature extraction. To cope with the problem of unavoidably ignorance in the integrity of objects caused by adopting the pixel-to-pixel attention strategy on flattened image features in existing transformer-based approaches, Song et al. ~\cite{29-song2022transformer} introduced multi-scale cyclic shifting window attention mechanism to transformer architecture, which expands the window samples with positional information and thus improves the accuracy.

In recent years, there has been a growing interest in a multi-modality tracking method called RGB-T tracking, which greatly enhances the accuracy and robustness of target tracking by combining information from both visible and infrared images. Accordingly, several typical RGB-T tracking methods have been presented as a supplement. The most prevalent RGB-T trackers are multi-domain network-based trackers. In order to jointly perform modality-shared, modality-specific and instance-aware target representation learning in the multi-domain network, Lu et al. ~\cite{lu2021rgbt} designed a multi-adapter network using the modified VGG-M and the hierarchical divergence loss to learn more shared features in the multi-adapter. Siamese network-based trackers are popular in RGB-T target tracking, e.g., Wang et al. ~\cite{wang2021multiple} proposed a four-stream oriented Siamese RGB-T tracker. By using co-attention mechanism for bilinear pooling and an inner product based logistical loss for training, this tracker successfully addressed the inevitable negative effects caused by the uninformed image blocks. Discriminative correlation filter is also an integral method of RGB-T target tracking. Zhang et al. ~\cite{zhang2022rgb} proposed a modality difference compensation module and a feature re-selection module, in order to reduce the modality differences between RGB and thermal features and then obtain the most discriminative features from both the unimodal features and the fused features.

\section{Method}\label{section:3}
In this section, we focus on the theory of our proposed TSF-SiamMU tracker. An overview of our tracker's framework is presented in section~\ref{section:3.1}; section~\ref{section:3.2} provides a detailed explanation of the twofold structured features network; on regard of the interference from appearance changes, section~\ref{section:3.3} thoroughly reports a multi-template update module based on template update mechanism; section~\ref{section:3.4} is a summary of the overall procedure of our tracking method.
\begin{figure*}
  \centering
{\includegraphics[width=16cm]{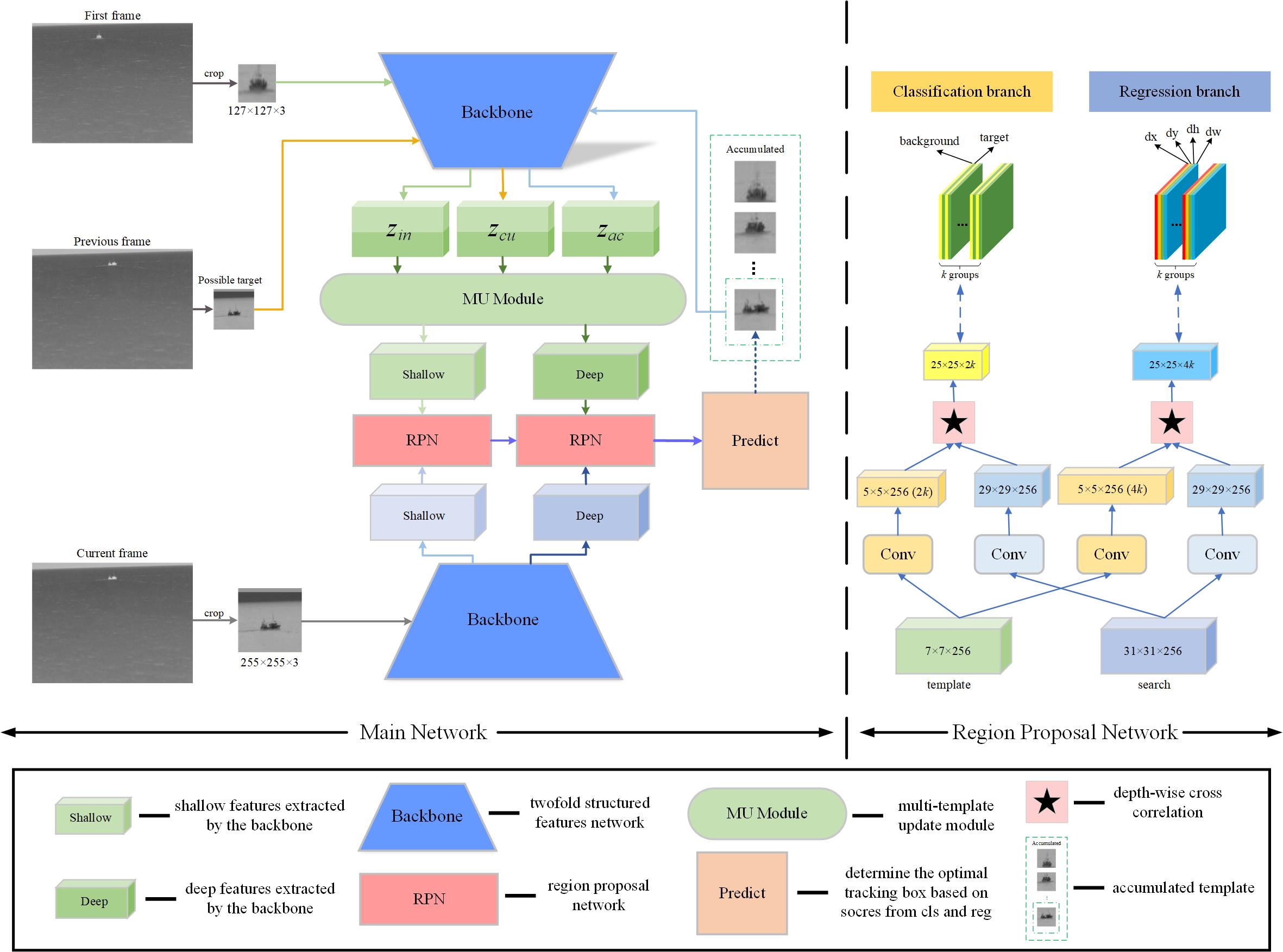}}
  \caption{The framework of the proposed TSF-SiamMU network. The left sub-figure shows its main structure, and the right one shows the structure of each RPN.}
\label{figure:1}
\end{figure*}
\subsection{Overview}\label{section:3.1}
Fig.~\ref{figure:1} shows the architecture of the proposed TSF-SiamMU network. The main network is composed of two parts: the backbone where the features are extracted and RPN ~\cite{22-li2018high} which performs bounding box prediction. The backbone, i.e., the twofold structured features network, is a kind of Siamese network consisting of a template branch and an instance branch. Two inputs into the two branches are the target image in the first frame and the cropped image in the current frame, i.e., template and instance. It is worth noting that “twofold” implies that the extracted output features in each branch are dual, which are respectively named as shallow features and deep features. While tracking, the optimal template will be changed into the weighted combination of all possible templates by MU module to adapt the tracker to appearance changes. When appearance changes of the target occur during tracking, MU module will output the optimal template of the target by weighting the combination of various templates.

The twofold structured features network, with both branches sharing almost the same structure and parameters, is the backbone part of the feature extraction network. In comparison to the instance branch, a MU module is additionally added on the template branch after the backbone outputs two distinguished layers of features, viz., shallow features and deep features. Note that although the backbone of the template branch seems to receive three inputs, they actually do not interfere each other and output three separate feature maps which contain both shallow features and deep features respectively.

The RPN consists of a classification branch and a regression branch, the former to handle the target-background classification responsible for identifying the target while the latter to calculate each candidate region through the regression values responsible for adjusting target orientation. Instead of up-channel cross correlation layers, depth-wise cross correlation layers are in the last part of RPN, in which the template features are convolved with the instance features to achieve sufficient information association and thus to generate a better output correspond map.

Therefore, when the Siamese backbone network and the RPN work together, it enables the single tracking process to be transformed into a one-shot detection task, for which the formula is given below:
\begin{equation}
\label{Eq:1}
\begin{split}
\mathop {\min }\limits_W \frac{1}{n}\sum\limits_{i = 1}^n L(\zeta (\varphi ({x_i};W);&{F_{MU}}(\varphi (z_i^{in};W),\\&
\varphi (z_i^{ac};W),\varphi (z_i^{cu};W))),{l_i})
\end{split}
\end{equation}
where $n$ represents the total number of data in one sequence; $x$, ${z_{in}}$, ${z_{ac}}$ and ${z_{cu}}$ denote the instance, the initial template, the accumulated template and the current template, respectively; $\zeta ( \cdot ; \cdot )$ and $\varphi ( \cdot ; \cdot )$ refer to the function of RPN and Siamese feature extraction network, accordingly; ${l_i}$ is the true candidate label in each frame; and all three different template features are fused by the function ${F_{MU}}( \cdot ; \cdot ; \cdot )$ to update the template, which will be discussed thoroughly in section~\ref{section:3.3}. To be more specific, it is argued that obtaining the parameter $W$ relies crucially on the average loss $L$.

\subsection{Twofold Structured Features Network }\label{section:3.2}
\begin{figure*}
  \centering
{\includegraphics[width=16cm]{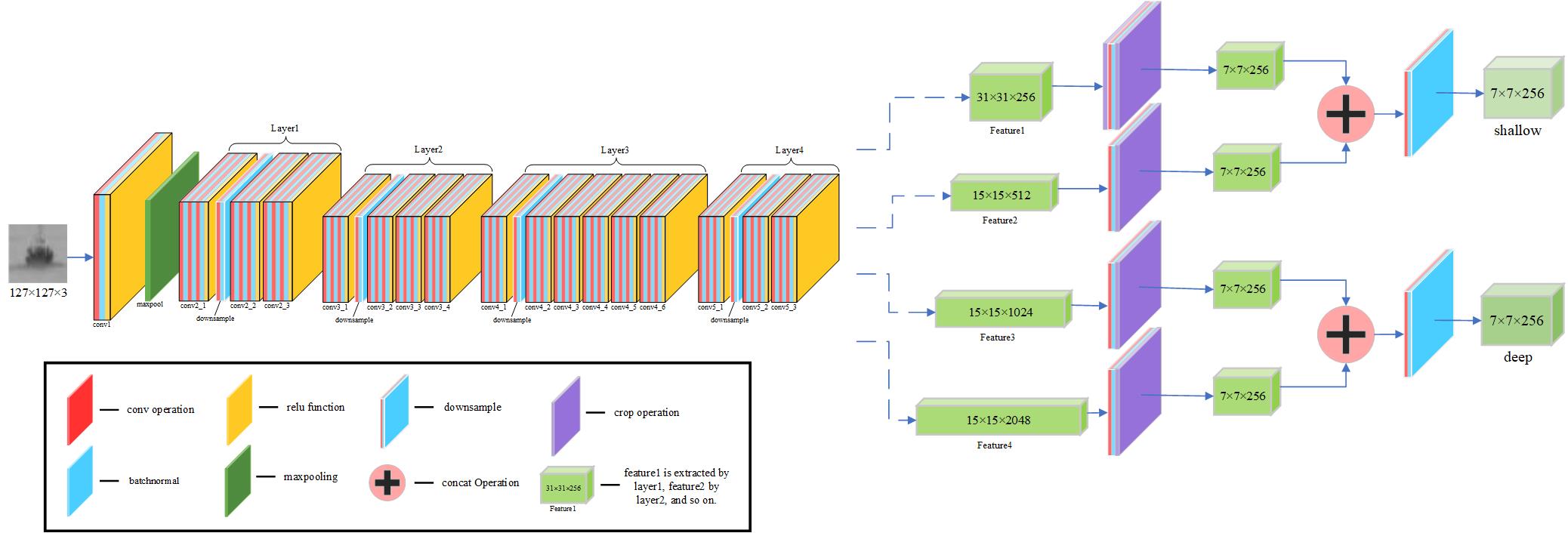}}
  \caption{Detailed structure of the proposed backbone network called twofold structured features network, where both shallow layers and deep layers are respectively fused to represent shallow and deep features of the target.}
\label{figure:2}
\end{figure*}

Inspired by the success of modern depth neural networks in Siamese trackers, we adopt ResNet-50 ~\cite{30-he2016deep} instead of AlexNet ~\cite{31-krizhevsky2012imagenet} as our base feature extraction network. To make the most of the feature representation capability of ResNet-50, we design a novel backbone network called twofold structured features network. The proposed network is depicted in Fig.~\ref{figure:2}, which consists of several layers and the subsequent adaptive fusion operation. We note that most existing Siamese trackers usually utilize features of last three layers to present the target, since these layers are at the same spatial resolution and easy to apply. However, as the network goes deeper, the extracted target features typically convey more semantic information than spatial information, which is prejudicial to infrared target tracking. Infrared images lack details such as texture and color, as a result of which the deep semantic information is far from discriminative when it comes to the infrared target, making spatial information more significant in infrared target tracking ~\cite{9-he2018twofold}. In addition, the extracted features of various layers are always a mixture of target information at different depth levels, yet we expect the features fed into the RPN to be more distinct and focused so as to better detect the target from the appearance and location separately. Given all these factors mentioned above, resolving the target robustly in complex infrared scenes can be a rather challenging task for the tracker.

In order to acquire more distinctive and robust features for the tracker, a relatively shallow layer, which is named as Layer1 in ResNet-50, is additionally employed in the proposed network as a supplement to the commonly used deeper layers, and all these layers are combined in a novel manner. Hence, the tracker is able to have access to a wealth of spatial information that facilitates the location of the infrared target. To make it clear, the sources of our shallow features and deep features are explained as a start. All layers are made up of several similar convolution combination and downsampling operation, both of which contain different types of convolution operations, normalization and ReLU activation function. Assume that $CN_{convkernel}^{outchannel}$ represents a single combination of convolution operation and batch normalization operation, in which the superscript refers to the output channel of the convolution while the subscript refers to the convolution kernel size, the structure of four layers for feature extraction can be formulated as the following formula:
\begin{equation}
\label{Eq:2}
\begin{aligned}
\left\{ \begin{array}{l}
layer1 = [CN_{1 \times 1}^{64};CN_{3 \times 3}^{64};CN_{1 \times 1}^{256}] \times 3\\
layer2 = [CN_{1 \times 1}^{128};CN_{3 \times 3}^{128};CN_{1 \times 1}^{512}] \times 4\\
layer3 = [CN_{1 \times 1}^{256};CN_{3 \times 3}^{256};CN_{1 \times 1}^{1024}] \times 6\\
layer4 = [CN_{1 \times 1}^{512};CN_{3 \times 3}^{512};CN_{1 \times 1}^{2048}] \times 3
\end{array} \right.
\end{aligned}
\end{equation}
\noindent where $n$ represents the number of sub-layers, and $[ \cdot ; \cdot ; \cdot ]$ denotes the structure of a single sub-layer. $[ \cdot ; \cdot ; \cdot ] \times n$ indicates the superposition of   sub-layers. 

Furthermore, we define ${F_1} \in {R^{31 \times 31 \times 256}}$, ${F_2} \in {R^{15 \times 15 \times 512}}$, ${F_3} \in {R^{15 \times 15 \times 1024}}$ and ${F_4} \in {R^{15 \times 15 \times 2048}}$ as the feature map extracted by Layer1, Layer2, Layer3 and Layer4, separately. As what we design, ${F_1}$ and ${F_2}$ carry more spatial information while ${F_3}$ and ${F_4}$ prevail in semantic information. In order to aggregate and then make sufficient use of the varied advantageous information in four feature maps, we need to apply several steps to capture and identify shallow features and deep features. Above all, we convert the four feature maps to the same spatial resolution and feature channels using function $CR( \cdot )$ and function $DS( \cdot )$, where $CR( \cdot )$ means crop operation to decrease the spatial resolution and $DS( \cdot )$ denotes downsampling operation to reduce the channel dimension. It is important to note that the initial spatial resolution of ${F_1}$ differs from other initial feature maps, as a result of which there is an extra crop operation for ${F_1}$ to apply before concatenation operation. After all the feature maps reach the size of $7 \times 7 \times 256$, we subsequently use a concatenation operation named $Concat( \cdot ; \cdot )$ to connect the feature maps sequentially along the channel direction. At last, another downsampling operation is performed to recover the feature map to the size of $7 \times 7 \times 256$, which eases the computational burden and still captures the entire target region. The whole process mentioned above can be expressed as follows:
\begin{equation}
\label{Eq:3}
\begin{split}
\left\{ \begin{array}{l}
{F_{shallow}} = DS(Concat(CR(DS(CR({F_1}))),\\\quad\quad\quad\quad\quad\quad\quad\quad\quad\quad\quad\quad\quad\quad CR(DS({F_2}))))\\
{F_{deep}} = DS(Concat(CR(DS({F_3})),\\\quad\quad\quad\quad\quad\quad\quad\quad\quad\quad\quad\quad\quad\quad CR(DS({F_4}))))
\end{array} \right.
\end{split}
\end{equation}
where ${F_{shallow}}$ represents the shallow feature map extracted by the backbone while ${F_{deep}}$ denotes the deep one. Either of them not only provides a decent representation of the target but is also the ideal size for both online tracking and offline training.

\subsection{Multi-template Update Module}\label{section:3.3}
\begin{figure}
  \centering
{\includegraphics[width=8.5cm]{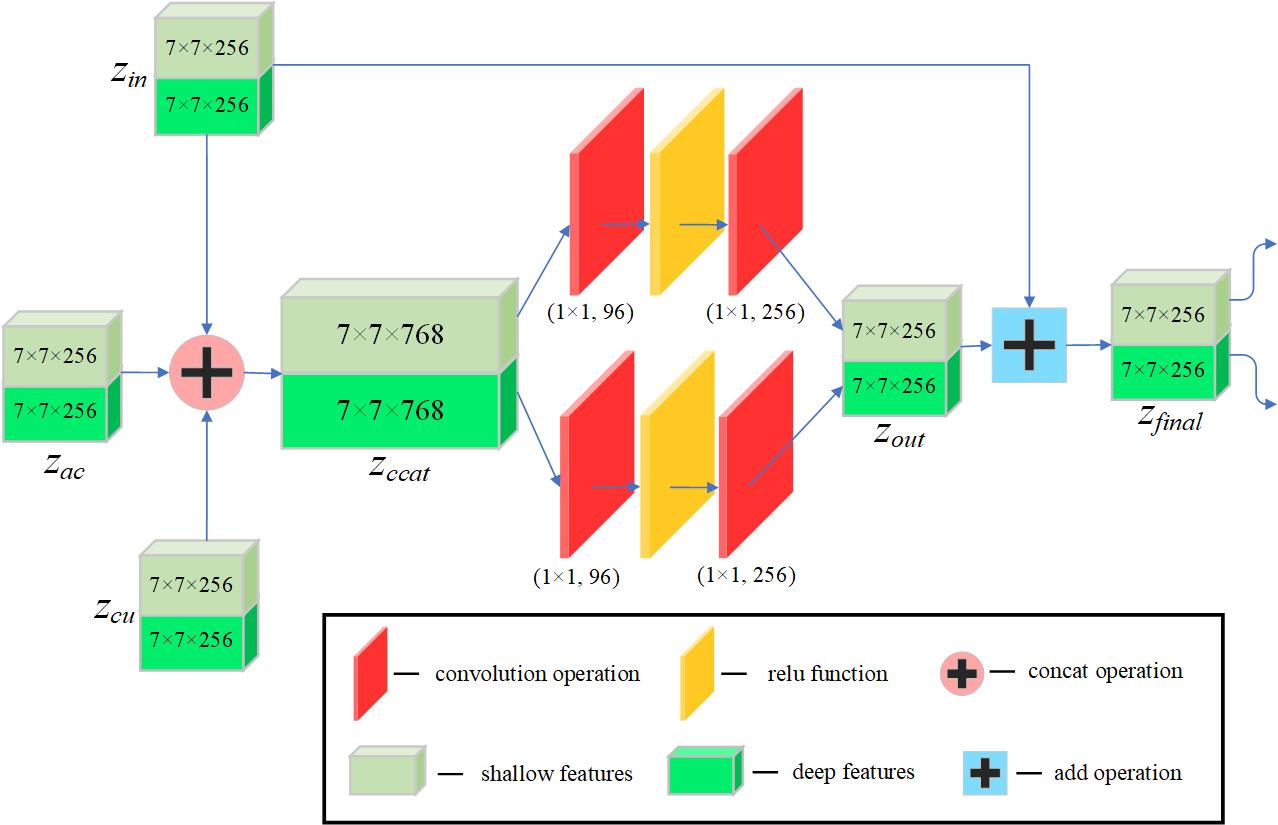}}
  \caption{Structure of the multi-template update module, in which initial template, accumulated template and current template are applied as supplements to the optimal template for target tracking.}
\label{figure:3}
\end{figure}
Due to large and frequent appearance changes, failing to update the template is a serious concern for conventional Siamese trackers. However, most linear update methods, where the template is updated as a running average with exponentially decaying weights over time, are too simple to allow for a flexible update mechanism when the infrared target undergoes complex appearance changes. Thus, we focus on designing a multi-template update module to obtain the optimal template suitable for tracking. 

In contrast to linear update methods, we adopt a simplified residual network to combine the templates in a far more reasonable manner throughout the tracking process. In the first step, the three different feature maps, all consisting of shallow and deep features, are connected sequentially along the channel direction. Once the concatenation operation is done, several convolution layers are applied to extract the detailed information in the accumulated template and the current template. It is necessary to point out that shallow features and deep features are handled in separate sub-networks with the same structure but varying parameters. Considering that the initial template provides the most reliable information, the final template is a combination of the initial template and all two output template extracted by the convolutional network. 

Since the target information in the current frame is rather important, the current template is extracted by the backbone based on the predicted position of the target in the previous frame. The final template is not only the optimal template to measure the bounding box for tracking in the current frame, but is also the accumulated template which we believe is accumulated with the information of all the past templates to calculate the optimal template in the next frame. The updating process is formulated as follows:
\begin{equation}\label{Eq:4}
\begin{split}
\left\{ \begin{array}{l}
{z_{final}^{T}} = MU_{}^{T}(Concat({z_{in}^{T}},{z_{ac}^{T}},{z_{cu}^{T}})) + {z_{in}^{T}}\\
MU_{}^{T} = CV_{}^{T}(RE_{}^{T}(CV_{}^{T}( \cdot )))\\
T = shallow,\ deep\\
{z_{final} = [{z_{final}^{shallow}}, {z_{final}^{deep}}]}
\end{array} \right.
\end{split}
\end{equation}
where ${z_{in}}$, ${z_{ac}}$ and ${z_{cu}}$ denote the initial ground-truth template, the last accumulated template and the current template extracted from the predicted target location, respectively. $CV( \cdot )$ represents the convolution operation while $RE( \cdot )$ represents the ReLU activation function. $Concat( \cdot ; \cdot ; \cdot )$ refers to the concatenation operation. $T$ denotes the different depth of the template operation, i.e. shallow or deep, and thus the templates of two depths perform a similar operation with varying parameters. The final template ${z_{final}}$ is likewise composed of shallow features and deep features.

\begin{table*}[ht]
\renewcommand{\arraystretch}{1.3}
  \centering
  \caption {ALGORITHMS FOR THE PROPOSED TSF-SIAMMU TRACKER.}
  \label{Table:1}
  \resizebox{\textwidth}{!}{
\begin{tabular}{l}
\hline
 \textbf{Algorithm 1} Main steps of the proposed TSF-SiamMU tracker\\
\hline
 \textbf{Input:} The first frame with the target ${z_0}$, initial bounding box $({x_0},{y_0},{w_0},{h_0})$, and the current frame ${x_i}$.\\
 \textbf{Output:} The predicted bounding box $({x_i},{y_i},{w_i},{h_i})$.\\
 \textbf{Repeat}\\
1. Crop the first frame ${z_0}$ based on the initial bounding box $({x_0},{y_0},{w_0},{h_0})$, extract it by Eq.~\ref{Eq:3}, and we get ${z_{in}}$.\\
2. Crop the current frame ${x_i}$, extract it by Eq.~\ref{Eq:3} the same as what we do to the template branch, and we get ${x_{cu}}$.\\
3. Estimate the optimal template ${z_{final}}$ via Eq.~\ref{Eq:4}, namely the Multi-template Update Module.\\
4. Calculate the $P_{cls}^{all}$ and $P_{reg}^{all}$ via Eq.~\ref{Eq:6}.\\
5. Calculate the highest scorer and its corresponding position via $P_{cls}^{all}$ and $P_{reg}^{all}$, and predict the bounding box $({x_i},{y_i},{w_i},{h_i})$.\\
 \textbf{Until} End of the video sequence.\\
\hline
\end{tabular}}
\end{table*}

The multi-template update module is trained to predict the target template $z_{i + 1}^{GT}$ which should be the best matching template to use when searching for the target in the next frame. Hence, to begin with, we extract the ground truth frame through the backbone (see Fig.~\ref{figure:2}) to obtain $z_{i + 1}^{GT}$. What is more, minimizing the Euclidean distance between the updated template and the ground-truth template is the key to derive $W$, just as what is defined below:
\begin{equation}\label{Eq:5}
\begin{split}
\mathop {\min }\limits_W {\left\| {{F_{MU}}(z_0^{in},z_i^{ac},z_i^{cu},W) - z_{i + 1}^{GT}} \right\|_2}
\end{split}
\end{equation}
where ${F_{MU}}( \cdot ; \cdot ; \cdot ;W)$ is used to update the template by fusing the three template features. $\left\|  \cdot  \right\|$ denotes the ${L_2}$ norm or Euclidean distance.

\subsection{General tracking process}
\label{section:3.4}
Once we obtain the search instance and the optimal template, the classification maps and the regression maps are naturally available via RPN (see Fig.~\ref{figure:1}). Since the shallow layers can capture fine-grained information useful for precise localization while the deep layers can encode abstract semantic information conducive to target recognition, we apply multiple adaptive prediction in order to take full advantage of multi-level features. We set the weights $\alpha$ and $\beta$ which correspond to each map and are capable to be optimized together with the backbone network. The formula can be expressed as follows:
\begin{equation}\label{Eq:6}
\begin{split}
\left\{ \begin{array}{l}
P_{cls}^{all} = {\alpha _s}P_{cls}^{shallow} + {\alpha _d}P_{cls}^{deep}\\
P_{reg}^{all} = {\beta _s}P_{reg}^{shallow} + {\beta _d}P_{reg}^{deep}
\end{array} \right.
\end{split}
\end{equation}
where ${P_{cls}}$ and ${P_{reg}}$ represent the classification maps and the regression maps, respectively. The classification branch scores each location calculated by the regression branch and the top scorer is the most likely to be the target location. 

To give a clear description of our proposed tracker, the main steps are summarized in Algorithm 1 (see Table~\ref{Table:1}).

\section{Experimental Results}\label{section:4}
In this section, we demonstrate the validity and sophistication of our proposed tracker through diverse experiments. First of all, the datasets we used to evaluate trackers are briefly outlined in section~\ref{section:4.1}. Then, we give some implementation details of offline and online acts in section~\ref{section:4.2}. Section~\ref{section:4.3} is an introduction of the evaluation criteria to quantify trackers’ performance. We carry out ablation experiments in section~\ref{section:4.4} in order to prove the effectiveness of each component in our tracker. In section~\ref{section:4.5}, other state-of-the-art trackers are compared with our tracker.

\subsection{Datasets}\label{section:4.1}
\begin{figure*}
  \centering
{\includegraphics[width=18cm]{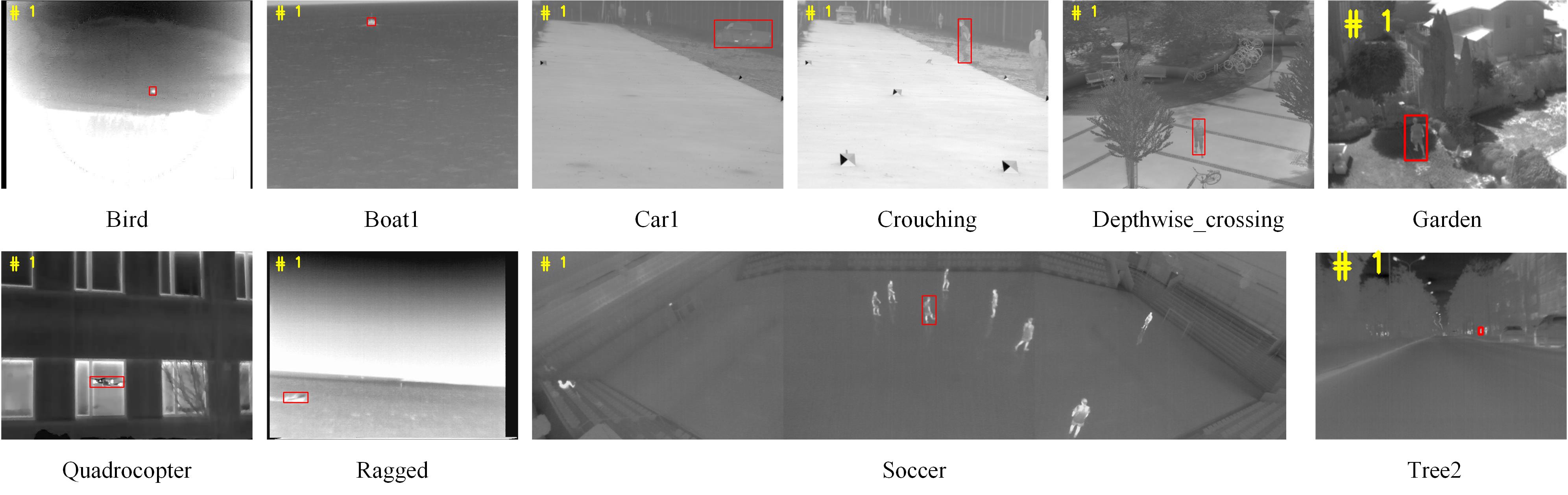}}
  \caption{First frame of part of the sequences in VOT-TIR 2016 dataset. The target is marked by a red rectangle.}
\label{figure:4}
\end{figure*}

The performance of our proposed method is measured using the VOT-TIR 2016 dataset ~\cite{32-lebeda2016thermal} in our experiments. The first frame of the partial sequence in VOT-TIR 2016 dataset is displayed in Fig.~\ref{figure:4}, with the red rectangle labelling the target of interest. Compared with VOT-TIR 2015 dataset ~\cite{33-felsberg2015thermal}, VOT-TIR 2016 dataset removes several sequences that are too easy for tracking and adds quite a number of challenging sequences, in which blur or motion change is a significant problem. In addition to the bounding box annotations, part of the local attributes and global attributes, i.e., size change, motion change, dynamic change, blur, scale variation, scene complexity and so on, are introduced in VOT-TIR 2016 dataset to present a better assessment of the comprehensive performance of the tracker.

In order to fully evaluate the performance of our proposed method, the GTOT dataset ~\cite{li2016learning} is used for additional comparison. The GTOT dataset contains a number of infrared-visible video sequences which are captured under various scenarios, including labs, campus roads, playgrounds, water pools, etc. All the corresponding groundtruth annotations are all done manually by the one person. Since we only discuss the infrared target tracking, it should be noted that our experiments are carried out on the thermal modality of the GTOT dataset.

\subsection{Implementation Details}\label{section:4.2}
Our experiments are conducted using PyTorch on a PC which is equipped with Intel i7-11700 CPU, NVIDIA GeForce RTX 3080Ti GPU and 16GB RAM. The proposed tracker can achieve an average running speed of 47 FPS.

We train our proposed backbone network with five large-scale datasets, including COCO ~\cite{34-lin2014microsoft}, GOT10k ~\cite{35-huang2019got}, LaSOT ~\cite{36-fan2019lasot}, ImageNet VID and ImageNet DET~\cite{37-russakovsky2015imagenet} in order to learn how to measure the similarities between objects for tracking. In both offline training and online tracking, we crop the frame in a fixed mode where the size of a template patch is $127 \times 127$ pixels and the size of an instance patch is $255 \times 255$ pixels. What is more, a stride-reduced ResNet-50 is applied as the pretrained network to initialise the parameters of the primary backbone network which will be further trained with stochastic gradient descent (SGD). Immediately thereafter, the training is implemented over 50 epochs where a warmup learning rate of 0.001 is used in the first 5 epochs and the learning rate exponentially decays from 0.005 to 0.0005 for the last 15 epochs. Noticing that the multi-template update module contains a neural network which is different from the common backbone network, we apply a unique training strategy in the next stage. Since the proposed multi-template update module has fewer parameters than the backbone network, we train this module with one single dataset named LaSOT to measure the optimal template for tracking. We train the module for 50 epochs with SGD and the learning rate is decreased logarithmically at each epoch from ${10^{ - 7}}$ to ${10^{ - 8}}$.

\subsection{Evaluation Criteria}\label{section:4.3}
To assess the performance of our presented tracker precisely, success rate and precision are adopted to evaluate the tracking performance. Above all, overlap score (OS) is defined for each frame in a sequence to represent the intersection rate of the predicted region and the ground-truth region, as well as center pixel error (PE) ~\cite{38-wang2015visual}~\cite{39-lu2014online} to represent the Euclidean distance in pixels between the predicted target center position and the ground-truth center position. The formulae of these two metrics are as follows:
\begin{equation}\label{Eq:7}
\begin{split}
OS = \frac{{\left| {{B_t} \cap {G_t}} \right|}}{{\left| {{B_t} \cup {G_t}} \right|}}
\end{split}
\end{equation}
\begin{equation}\label{Eq:8}
\begin{split}
PE = \sqrt {{{\left( {x_t^B - x_t^G} \right)}^2} + {{\left( {y_t^B - y_t^G} \right)}^2}}
\end{split}
\end{equation}
where ${B_t}$ and ${G_t}$ denote the predicted bounding box and the real location of the target, correspondingly. The coordinates of the centers of ${B_t}$ and ${G_t}$ are denoted as $\left( {x_t^B,y_t^B} \right)$ and $\left( {x_t^G,y_t^G} \right)$, respectively.

Then, with the thresholds $O{S_{th}}$ and $P{E_{th}}$ configured, the success rate ${S_i}$ and the precision rate ${P_i}$ in a given sequence can be calculated as the ratio of frames ${k_i}$ above or below the threshold to the total number of frames ${n_i}$. The definitions are outlined below:
\begin{equation}\label{Eq:9}
\begin{split}
{S_i} = \frac{{{k_{OS > O{S_{th}}}}}}{{{n_i}}}
\end{split}
\end{equation}
\begin{equation}\label{Eq:10}
\begin{split}
{P_i} = \frac{{{k_{PE < P{E_{th}}}}}}{{{n_i}}}
\end{split}
\end{equation}
where $k$ denotes the specific frames selected from all the frames. As we set the threshold $O{S_{th}}$, we obtain all the frames whose $OS$ values are larger than $O{S_{th}}$, and we defined it as ${k_{OS > O{S_{th}}}}$. Subsequently, by setting different $O{S_{th}}$ thresholds, the success rates are obtained and a success plot is thus formed. Similarly, ${k_{PE < P{E_{th}}}}$ can be defined and in the same way, a precision plot is able to be drawn. We typically report the area under curve in the success plot to represent the comprehensive capability of the tracker in tracking success rates, whereas the precision at threshold $P{E_{th}}$ of 20 pixels in the precision plot is usually reported as the representative precision of the tracker.

\begin{figure}[ht]
  \centering
{\includegraphics[width=8.5cm]{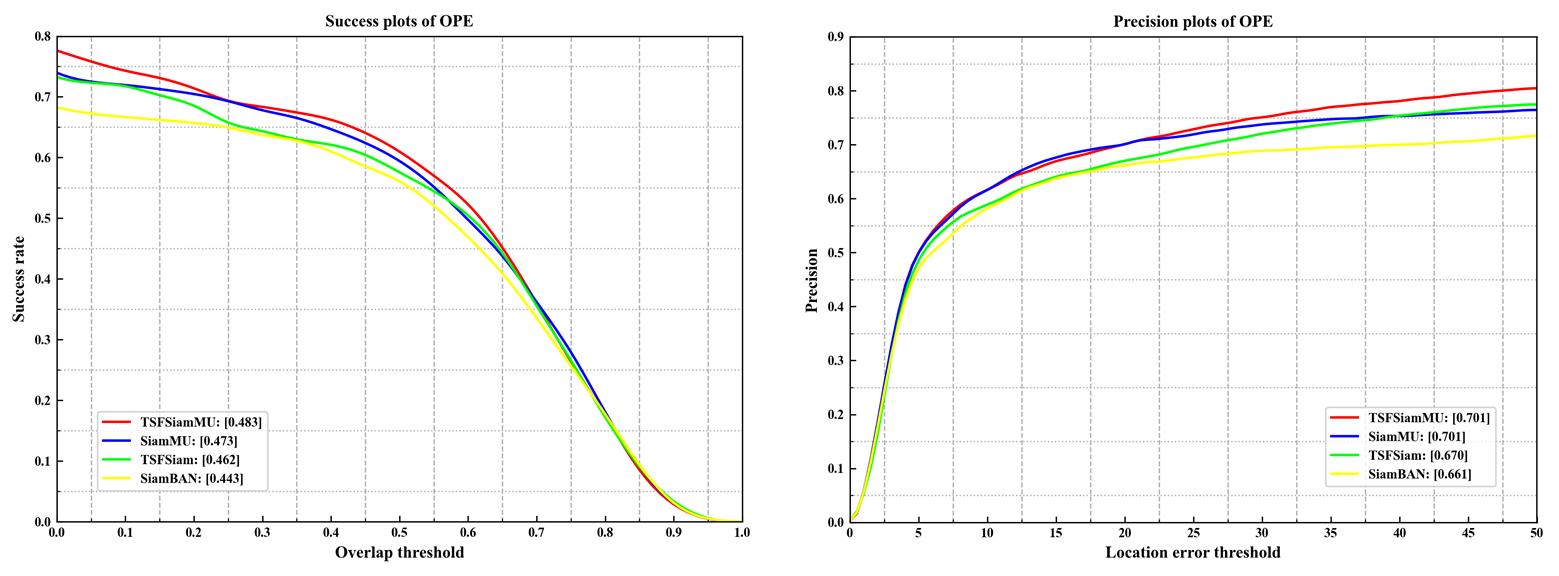}}
  \caption{Success plots and precision plots of the ablation experiments conducted on VOT-TIR 2016 dataset.}
\label{figure:5}
\end{figure}
\begin{table}
\renewcommand{\arraystretch}{1.3}
  \centering
  \caption{ABLATION ANALYSES OF OUR METHOD ON VOT-TIR 2016 DATASET}
  \label{Table:2}
\begin{tabular}{c|c|c c c}
\hline
Component&SiamBAN&TSF-Siam&SiamMU&TSF-SiamMU\\
\hline
TSF& &\checkmark& &\checkmark\\
MU& & &\checkmark&\checkmark\\
\hline
AUC&0.443&0.462&0.473&0.483\\
Precision&0.661&0.670&0.701&0.701\\
\hline
\multicolumn{5}{l}{\makecell[l]{TSF: twofold structured features network, MU: multi-template update\\ module.}}
\end{tabular}
\end{table}
\begin{figure*}[ht]
  \centering
{\includegraphics[width=17cm]{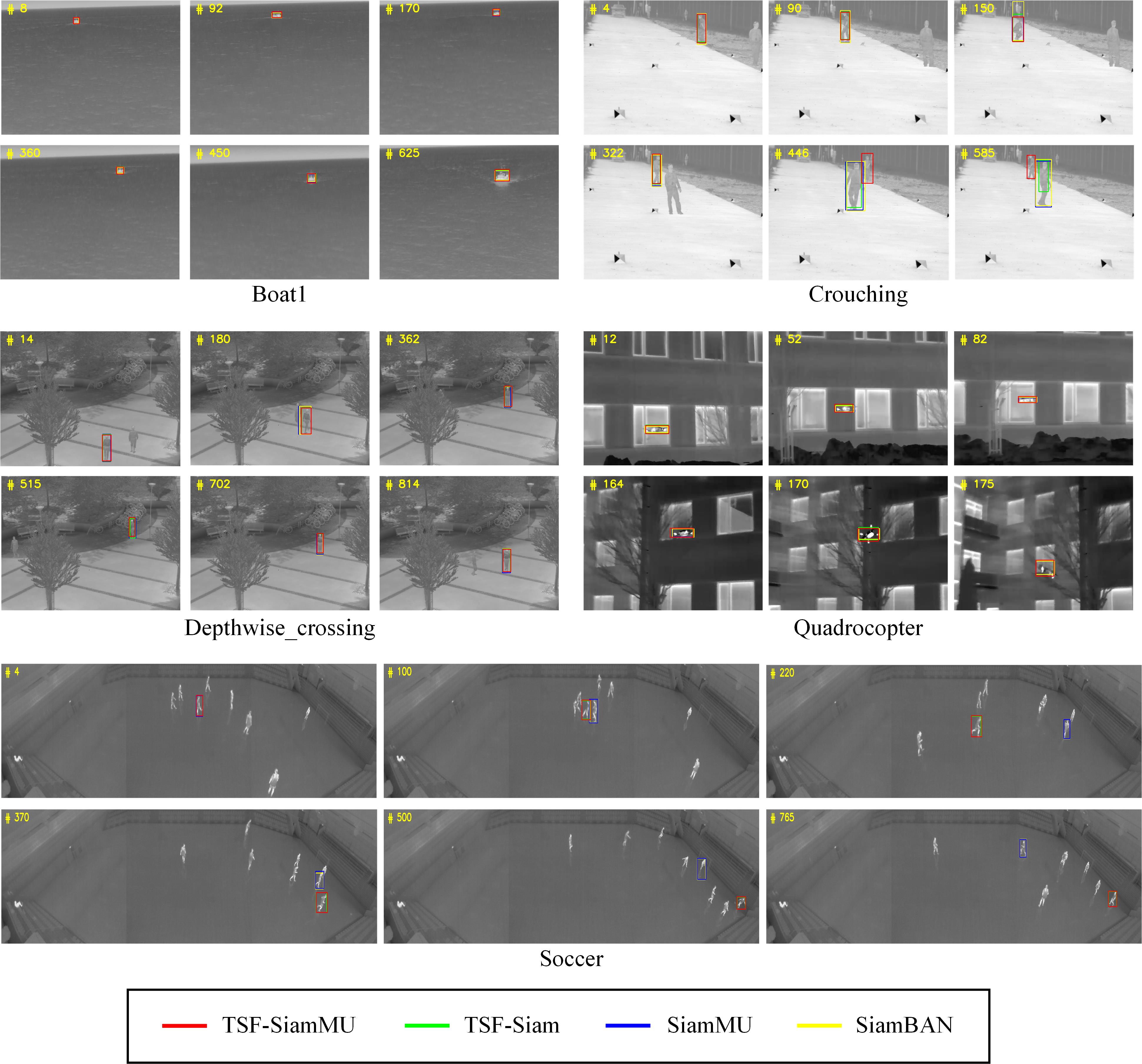}}
  \caption{Part of the visualization results of the ablation experiments conducted on VOT-TIR 2016 dataset.}
\label{figure:6}
\end{figure*}
\begin{figure}[ht]
  \centering
{\includegraphics[width=8.5cm]{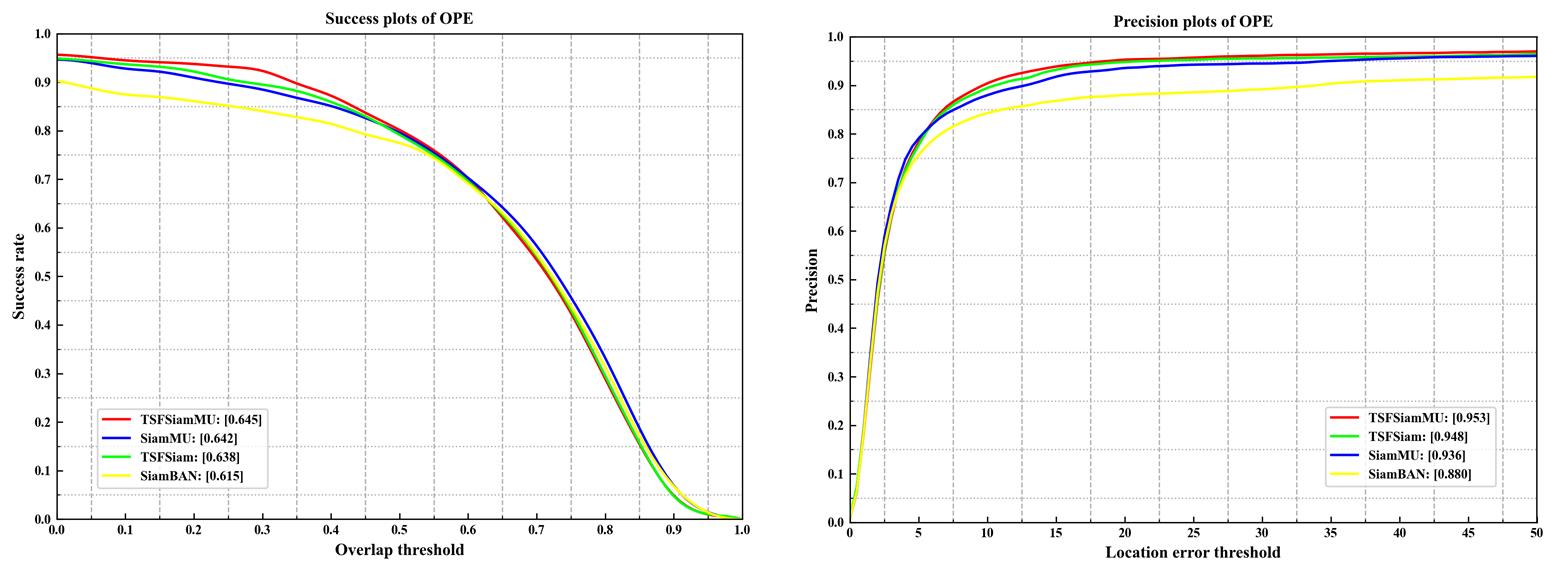}}
  \caption{Success plots and precision plots of the ablation experiments conducted on GTOT dataset.}
\label{figure:7}
\end{figure}
\begin{table}
\renewcommand{\arraystretch}{1.3}
  \centering
  \caption{ABLATION ANALYSES OF OUR METHOD ON GTOT DATASET}
  \label{Table:3}
\begin{tabular}{c|c|c c c}
\hline
Component&SiamBAN&TSF-Siam&SiamMU&TSF-SiamMU\\
\hline
TSF& &\checkmark& &\checkmark\\
MU& & &\checkmark&\checkmark\\
\hline
AUC&0.615&0.638&0.642&0.645\\
Precision&0.880&0.948&0.936&0.953\\
\hline
\multicolumn{5}{l}{\makecell[l]{TSF: twofold structured features network, MU: multi-template update\\ module.}}
\end{tabular}
\end{table}
\subsection{Ablation Analysis}\label{section:4.4}
In this section, we verify the validity of each optimization component in our tracker through ablation analysis. We carry out an internal comparison experiment on VOT-TIR 2016 and GTOT datasets to evaluate the three variants of our method.

Since all modified components are added on the basis of the baseline tracker called SiamBAN, we compare our variants with SiamBAN and our proposed TSF-SiamMU tracker to prove that each of our improvements is generally positive. First, a tracker named TSF-Siam, which only applies the twofold structured features network, is compared with SiamBAN to check whether our proposed twofold structured features network is effective. Subsequently, in a bid to prove the multi-template update module works, SiamMU which is only equipped with the multi-template update module is compared with SiamBAN. Finally, a comprehensive comparison of TSF-Siam, SiamMU and TSF-SiamMU is performed in order to demonstrate that all of our optimization components are uncontradictory and synergistic once they are combined all together, where TSF-SiamMU is our proposed tracker. 

The success plots and precision plots of one-path evaluation (OPE) ~\cite{40-wu2013online} on VOT-TIR 2016 dataset is depicted in Fig.~\ref{figure:5}. Moreover, Table~\ref{Table:2} are tabular summarised results of ablation experiments on VOT-TIR 2016 dataset. From the graph and the table, it is apparent that every optimization component included in our tracker makes a considerable contribution to improving the tracking performance. To begin with, it is plain to witness that TSF-Siam outperforms the baseline algorithm SiamBAN not only in terms of AUC but also in terms of precision. Especially, the AUC and precision are almost 4.3\% and 1.4\% higher than SiamBAN, which demonstrates that the twofold structured features network facilitates the extraction of more robust and distinctive infrared target features. Next, when adopting the multi-template update module to the baseline algorithm, the AUC and precision of SiamMU are increased by 6.8\% and 6.1\% compared to SiamBAN, which proves that the multi-template update module is a valid tool for the target tracking task. Last but not least, although the precision is almost equal for TSF-SiamMU and SiamMU, when it comes to AUC, TSF-SiamMU gains 4.5\% and 2.1\% increase compared with TSF-Siam and SiamMU, respectively, which implies that the two components cooperate well with each other.

\begin{table*}
\renewcommand{\arraystretch}{1.3}
  \centering
  \caption{COMPARISON BETWEEN TSF-SIAMMU AND OTHER STATE-OF-THE-ART TRACKERS ON VOT-TIR 2016 dataset. DATA MARKED IN RED, GREEN AND BLUE INDICATES THE FIRST, SECOND, AND THIRD BEST PERFORMANCE, RESPECTIVELY.}
  \label{Table:4}
  \resizebox{\textwidth}{!}{
\begin{tabular}{c|c c c c c c c c c c c c}
\hline
Tracker&TSF-SiamMU&SiamCAR&TCTrackerpp&CSWinTT&SiamRPNpp&UpdateNet&SiamBAN&SiamMask&SiamFC&MCCTH&STRCF&DCF\\
\hline
AUC&{\bf \color{green} 0.483}&{\bf \color{blue}  0.467}&0.434&{\bf \color{red}  0.494}&0.458&0.427&0.443&0.402&0.343&0.446&0.359&0.315\\
Precision&{\bf \color{red} 0.701}&{\bf \color{blue} 0.662}&0.640&{\bf \color{green} 0.680}&0.652&0.619&0.661&0.614&0.519&0.631&0.558&0.500\\
FPS&47.4&71.1&{\bf \color{blue} 99.1}&9.5&76.8&{\bf \color{green} 118.5}&80.0&65.8&71.9&21.4&24.1&{\bf \color{red} 509.7}\\
\hline
\end{tabular}}
\end{table*}
\begin{figure}
  \centering
{\includegraphics[width=8.5cm]{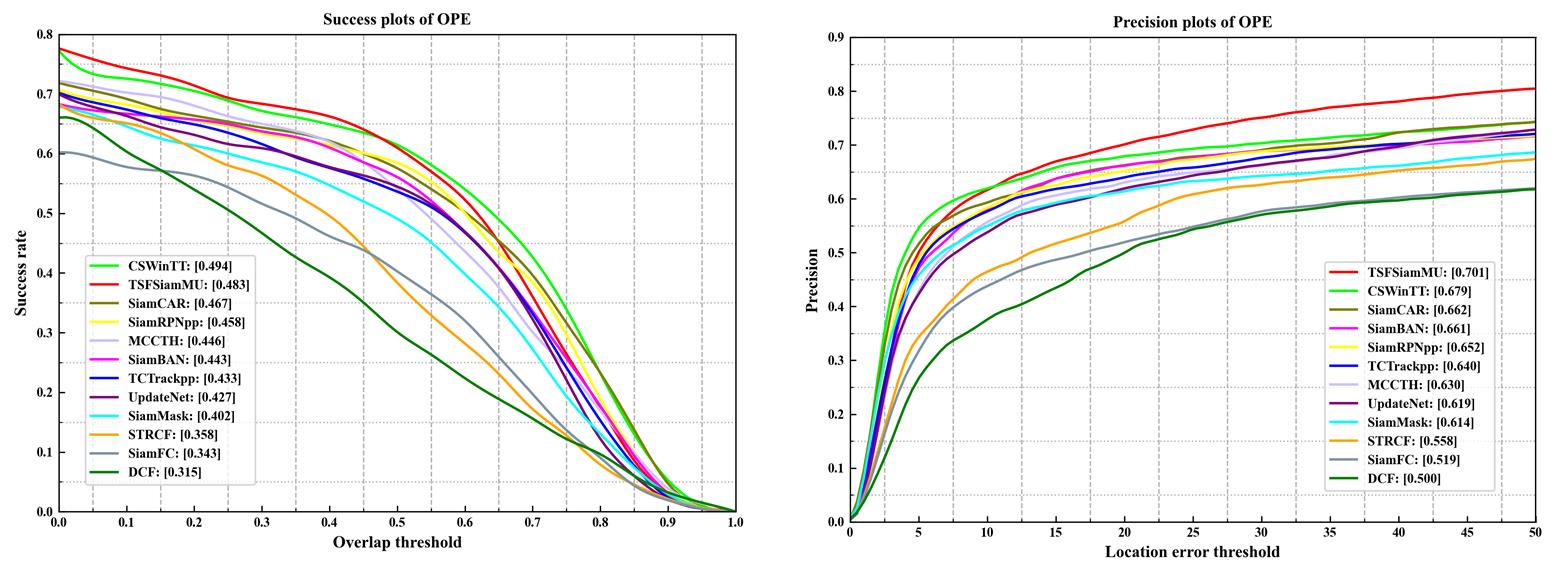}}
  \caption{Success plots and precision plots of the external comparison experiments conducted on VOT-TIR 2016 dataset.}
\label{figure:8}
\end{figure}
As shown in Fig.~\ref{figure:6}, for the purpose of demonstrating the performance of our three variants (TSF-Siam, SiamMU and TSF-SiamMU) and the baseline tracker (SiamBAN) in a more intuitive manner, we select several challenging sequences from VOT-TIR 2016 dataset and visualize the results on these sequences. It is evident that the proposed TSF-SiamMU is able to cope with complex target deformation and consistently localise the target. In some sequences (see “Crouching” and “Soccer”), once the target changes in size, TSF-Siam is prone to tracking drift when it encounters the interference caused by similar deep semantic information in the following tracking. Meanwhile, without the ability of meticulous discrimination between shallow and deep information and thus lack of robust template features, SiamMU is vulnerable to those confusing similar targets in close proximity to each other, which brings instability to the results of infrared tracking. Finally, equipped with twofold structured features network and multi-template update module, TSF-SiamMU not only runs in a more stable way (see sequence “Boat1”, “Depthwise\_crossing” and “Quadrocopter”) but also more robust to those complex tracking scenes (see sequence “Crouching” and “Soccer”).

We also perform supplementary experiments on GTOT dataset to further evaluate the three variants of our method. The success plots and precision plots of one-path (OPE) evaluation on GTOT dataset are depicted in Fig. ~\ref{figure:7} and the detailed results of experiments are presented in Table ~\ref{Table:3}. Both the graph and the table demonstrate that each optimization component included in our tracker does contribute to the tracking performance when handling infrared images. Above all, TSF-SiamMU outperforms the baseline algorithm SiamBAN not only in terms of AUC but also in terms of precision, namely, the AUC and precision of TSF-SiamMU are 4.9\% and 8.3\% higher than those of SiamBAN, respectively. In particular, equipped with twofold structured features network or multi-template update module, the performance of both TSF-Siam and SiamMU considerably exceeds the baseline SiamBAN. Nevertheless, the AUC of TSF-Siam is still 1.1\% lower than that of TSF-SiamMU, and the precision of SiamMU is 1.8\% lower than that of TSF-SiamMU. In conclusion, when dealing with infrared images, two components cooperate well with each other and are all valid tools for the infrared tracking task.

\begin{figure*}
  \centering
{\includegraphics[width=17cm]{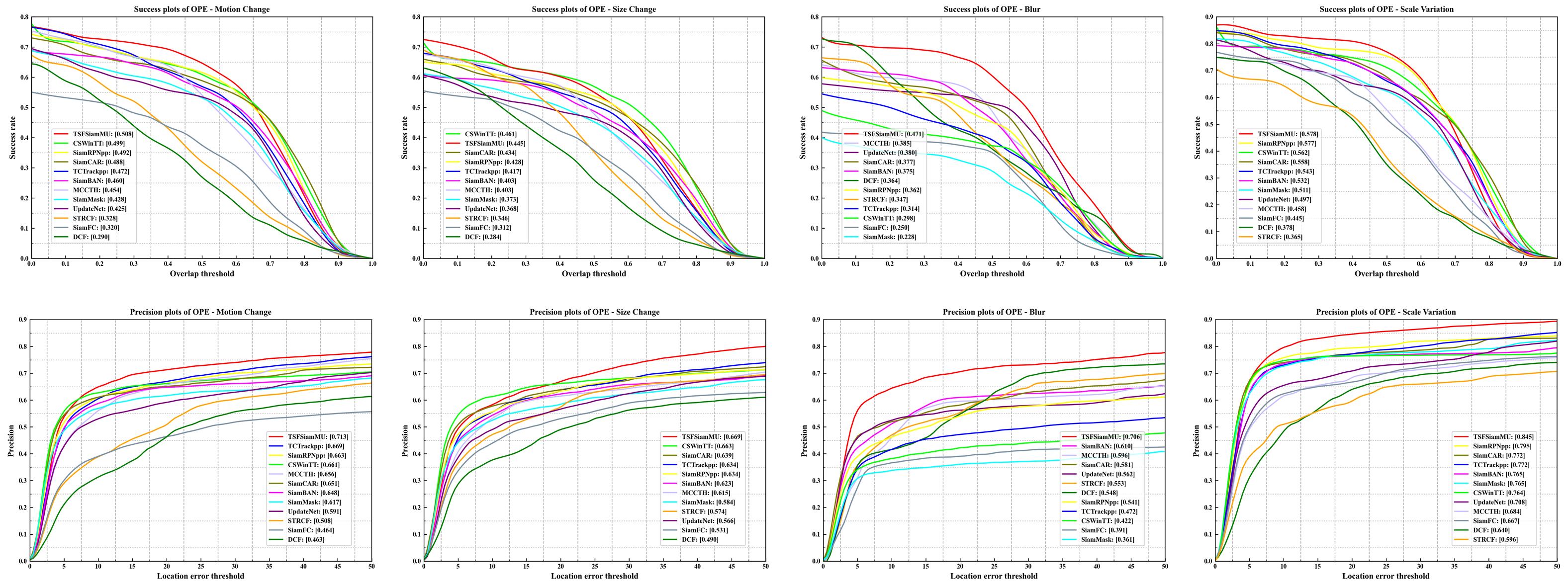}}
  \caption{Success plots and precision plots of the external comparison experiments conducted on sequences with different challenges.}
\label{figure:9}
\end{figure*}
\begin{figure*}
  \centering
{\includegraphics[width=17cm]{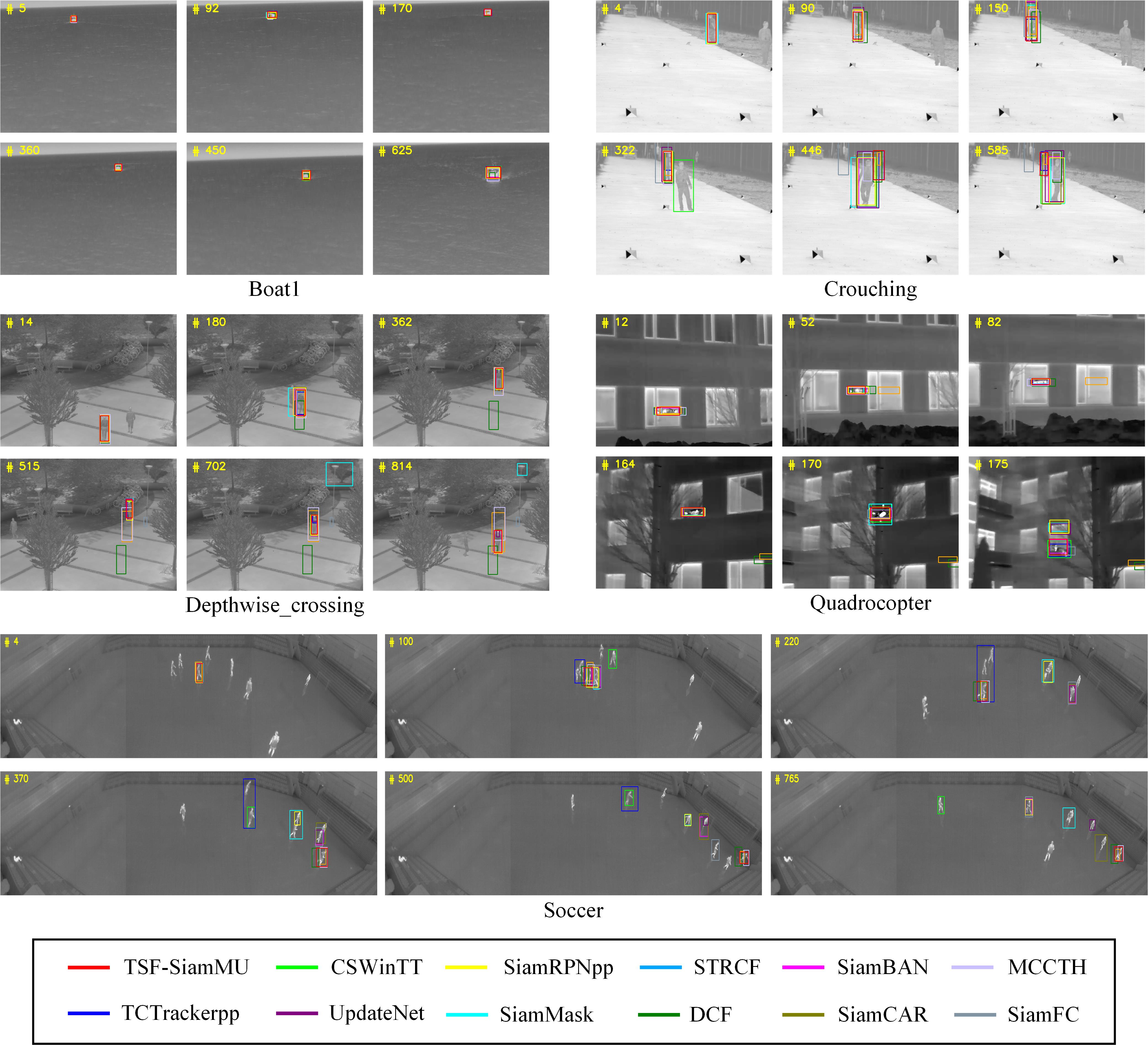}}
  \caption{Part of the visualization results of the external comparison experiments conducted on VOT-TIR 2016 dataset.}
\label{figure:10}
\end{figure*}
\subsection{External Comparison}\label{section:4.5}
In this section, in order to demonstrate that our tracker obtains the stronger performance on VOT-TIR 2016 dataset against the state-of-the-art algorithms, we compare our method with another 11 different tracking algorithms. The 11 tracking algorithms are grouped into three categories: Siamese trackers consisting of SiamFC ~\cite{7-bertinetto2016fully}, SiamMask ~\cite{43-wang2019fast}, UpdateNet ~\cite{10-zhang2019learning}, SiamRPNpp ~\cite{8-li2019siamrpn++}, SiamCAR ~\cite{44-guo2020siamcar}, SiamBAN ~\cite{23-chen2020siamese} and TCTrackerpp ~\cite{6-cao2022tctrack}; CF trackers including DCF ~\cite{15-henriques2014high}, STRCF ~\cite{41-li2018learning} and MCCTH ~\cite{42-wang2018multi}; and one transformer tracker CSWinTT ~\cite{29-song2022transformer}. 

\begin{table*}
\renewcommand{\arraystretch}{1.3}
  \centering
  \caption{ADDITIONAL COMPARISON BETWEEN TSF-SIAMMU AND STATE-OF-THE-ART TRACKERS ON GTOT DATASET. DATA MARKED IN RED, GREEN AND BLUE INDICATES THE FIRST, SECOND, AND THIRD BEST PERFORMANCE, RESPECTIVELY.}
  \label{Table:5}
  \resizebox{\textwidth}{!}{
\begin{tabular}{c|c c c c c c c c c c c c}
\hline
Tracker&TSF-SiamMU&SiamCAR&TCTrackerpp&CSWinTT&SiamRPNpp&UpdateNet&SiamBAN&SiamMask&SiamFC&MCCTH&STRCF&DCF\\
\hline
AUC&{\bf \color{red} 0.645}&{\bf \color{blue}  0.636}&0.561&0.622&{\bf \color{green}  0.638}&0.619&0.615&0.572&0.538&0.575&0.552&0.482\\
Precision&{\bf \color{red} 0.953}&{\bf \color{green} 0.927}&0.812&0.834&0.915&0.898&0.880&{\bf \color{blue} 0.916}&0.847&0.842&0.818&0.798\\
FPS&45.2&79.7&{\bf \color{blue} 140.2}&11.4&81.5&131.0&76.8&90.6&{\bf \color{green} 175.4}&24.7&25.7&{\bf \color{red} 905.2}\\
\hline
\end{tabular}}
\end{table*}
\begin{figure}
  \centering
{\includegraphics[width=8.5cm]{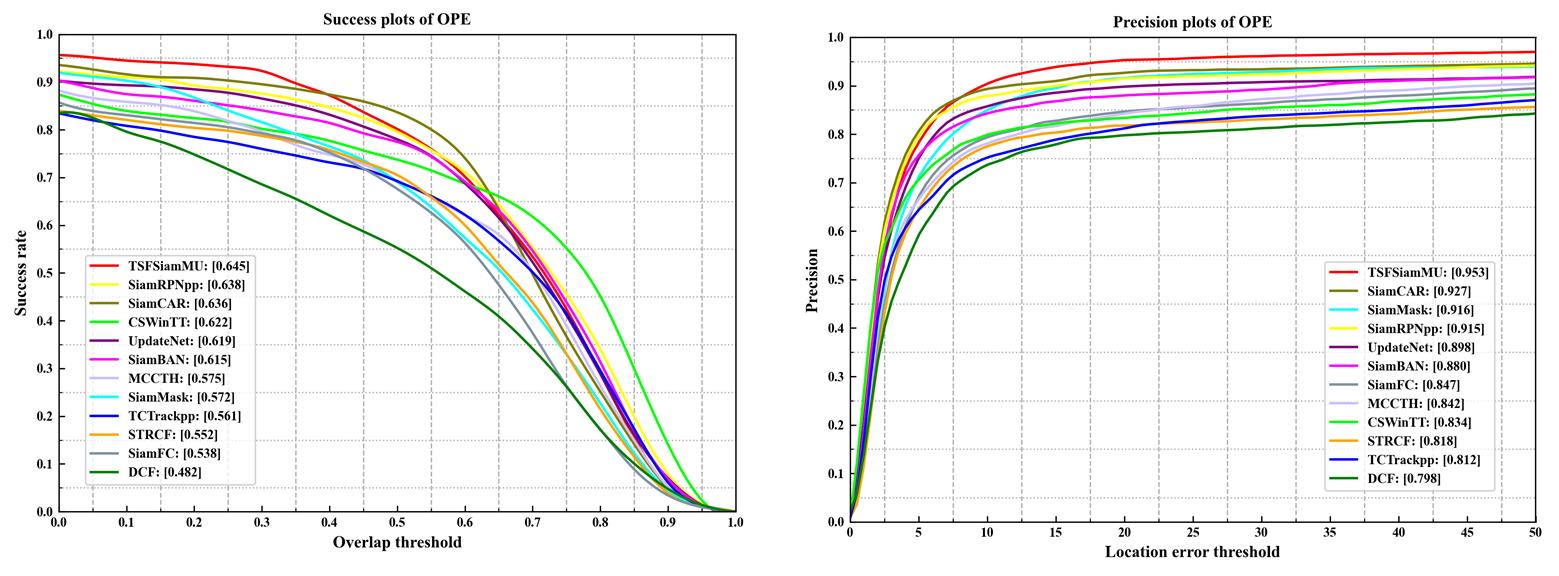}}
  \caption{Success plots and precision plots of the additional comparison experiments conducted on GTOT dataset.}
\label{figure:11}
\end{figure}
The success plots and precision plots of OPE on VOT-TIR 2016 dataset are depicted in Fig.~\ref{figure:8}. All detailed test results are presented in Table~\ref{Table:4}. It should be noted that all parameter values of the trackers utilized for comparison are default parameters set or trained by their own authors. First of all, our TSF-SiamMU tracker attains the second AUC of 0.483 amongst all these methods, which is only 2.2\% lower than that of the non-real-time tracker CSWinTT and 3.4\% higher than that of the real-time tracker SiamCAR. Since CSWinTT runs at 9.5 FPS far behind the real-time requirement of 30 FPS, it is acceptable that our method almost catches up with CSWinTT which obtains the AUC of 0.494. CF trackers fall far behind our method obviously, where the best CF tracker MCCTH acquires the AUC of 0.446. SiamCAR and SiamRPNpp achieve 0.467 and 0.458 in terms of AUC, respectively, which are the top two Siamese trackers for comparison but still not as strong as our method. In addition, it is significant to note that our tracker achieves the Precision of 0.701, which comes first among all mentioned trackers even including CSWinTT. For remaining Siamese trackers, SiamCAR stands out with its superior precision of 0.662, yet having a 5.9\% decrease when compared with our method. Although the CF tracker MCCTH reaches the precision of 0.631, it is still 11.1\% lower than ours, which further implies the advance of our method in accuracy. Moreover, the third rows in Table~\ref{Table:4} show the exact results of tracking speed. Our proposed TSF-SiamMU tracker achieves a speed of 47.4 FPS, satisfying the requirement of real-time tracking. Though DCF gains a speed of more than 500 FPS, its tracking performance falls far behind ours. Whereas, these trackers for comparison either perform the equivalent performance to TSF-SiamMU but hardly meet real-time requirements, as in the case of CSWinTT; or they meet real-time requirements yet are nowhere near as well as ours in terms of tracking performance, as in the case of SiamCAR and SiamRPNpp. In conclusion, we believe that our tracker strikes a balance between tracking performance and tracking speed, emerging as a new outstanding tracker for infrared target tracking.

In order to analyse the varying influence of different tracking challenges on our trackers in VOT-TIR 2016 dataset, certain groups of sequences with challenges of motion change, size change, blur and scale variation are selected for comparison experiments. All evaluation sequences in these four groups are chosen from VOT-TIR 2016 dataset on the basis of the given attributes. Fig.~\ref{figure:9} shows the detailed comparison results. To begin with, when encountering objects with motion changes, our tracker is the best tracker with respect to both AUC and precision, owing to the design of our feature extraction network which separates shallow features and deep features. In contrast, even though CSWinTT has strong feature extraction capability at the expense of tracking speed, it makes mistakes from time to time due to its lack of distinction between spatial and semantic information of the target, thus resulting in the inferior performance to TSF-SiamMU in sequences of motion changes. Second, due to the inclusiveness to size changes in targets by using multi-template update module, our method takes the second place on AUC and the first place on precision. Relatively, trackers which utilize nothing or simple linear template update strategy, such as SiamCAR and SiamBAN, is sensitive to the appearance changes of target, especially complex shape changes. Third, it is plain to see that our TSF-SiamMU tracker gets the highest AUC and percision when it comes to those blur sequences. Despite the difficulty of these sequences and the fact that none of the trackers get the satisfied performance, our method makes a step forward nevertheless thanks to its comprehensive ability of deriving the semantic information from the background. Besides, we can see clearly that either the AUC or the precision of TSF-SiamMU ranks first in sequences with frequent scale variation. Overall, judging from the curves of different challengings, TSF-SiamMU does have the outstanding performance.

The outstanding performance of our TSF-SiamMU tracker goes far beyond the figures, which can also be found from the selected visualization results shown in Fig.~\ref{figure:10}. When the target comes across motion changes, many trackers, including Siamese trackers, suffer the risk of tracking drift. In particular, it is worsened if the infrared target encounters occlusion, blur or size changes during movement. The tracking results in sequences “Crouching”, “Depthwise\_crossing” and “Soccer” are able to support this conclusion. In terms of sequences with scale variation challenges, such as “Boat1” and “Quadrocopter”, the changes in target appearance information are coherent and continuous, so the tracking accuracy must be maintained at a high level throughout the tracking process. In this regard, TSF-SiamMU not only deploys twofold structured features network to make full use of the shallow and deep features of the target, but also applies multi-template update module to attenuate the interference of target appearance changes. All in all, both quantitative and visualization results indicate that our TSF-SiamMU tracker is capable of standing up to other state-of-the-art methods in real infrared scenes.

\begin{figure*}
  \centering
{\includegraphics[width=17cm]{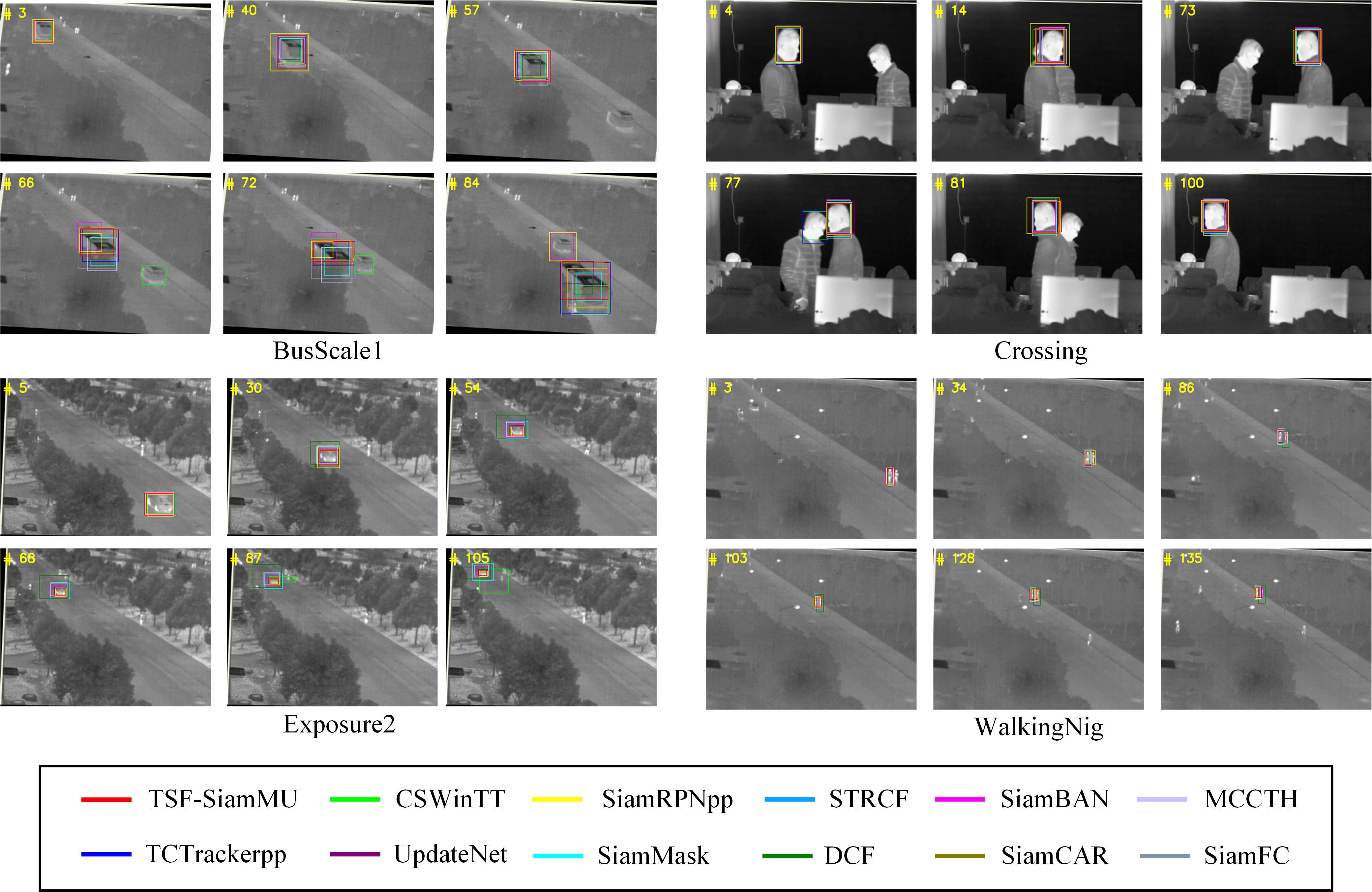}}
  \caption{Part of the visualization results of the additional comparison experiments conducted on GTOT dataset.}
\label{figure:12}
\end{figure*}

\subsection{Additional Comparison}\label{section:4.6}
In this section, we conduct some additional comparison experiments to demonstrate that our tracker remains the promising performance not only on VOT-TIR 2016 dataset, but also on other TIR datasets, e.g., GTOT, compared to other 11 trackers algorithms mentioned in section~\ref{section:4.5}. It is worth noting that since we only discuss infrared target tracking, our experiments are implemented on the thermal modality of GTOT dataset.

The success plots and precision plots of OPE on GTOT dataset are depicted in Fig.~\ref{figure:11}. All detailed test results are presented in Table~\ref{Table:5}. Note that all parameter values of the trackers utilized for comparison are default parameters set or trained by their own authors. Among these methods, our TSF-SiamMU tracker achieves the best AUC of 0.645 on GTOT, which is 1.1\% and 1.4\% higher than that of SiamRPNpp and SiamCAR. In contrast, the remaining Siamese trackers fall far behind our method with the lowest AUC of 0.538. Among all CF trackers, MCCTH obtains the best AUC of 0.575, yet still lags far behind our method. The transformer tracker CSWinTT attains the AUC of 0.622 on GTOT, 3.7\% lower than that of our method. What is more, it is notable that our method outperforms all other 11 trackers with the Precision of 0.953, i.e., 2.8\% higher than that of SiamCAR which gains the second Precision of 0.927. For remaining Siamese trackers, SiamMask and SiamRPNpp reach the Precision of 0.916 and 0.915 respectively, i.e., about 4\% lower than that of our method. Although the CF tracker MCCTH attains the decent Precision of 0.842, it is still approximately 13\% lower than ours. CSWinTT only obtains the Precision of 0.834, which is approximately 12.5\% lower than our method. Additionally, the third rows in Table~\ref{Table:5} show the average tracking speed of each tracker. Our proposed TSF-SiamMU tracker achieves a speed of 45.2 FPS on average, satisfying the requirement of real-time tracking. In general, our TSF-SiamMU tracker outperforms other trackers with similar tracking speed, as in the case of SiamCAR and SiamRPNpp. Although DCF runs at nearly 905 FPS, its tracking performance is much poorer than our tracker.

In order to visualize the outstanding performance of our TSF-SiamMU tracker, part of the visualization results on GTOT dataset are shown in Fig.~\ref{figure:12}. As we can see, many trackers suffer from the risk of tracking drift when the infrared target is not well characterized or even occluded in some sequences such as “Crossing” and “WalkingNig”. Therefore, richer and more robust feature representations are needed to overcome the difficulties in these sequences, where our tracker performs better for the benefit of our novel design of the twofold structured features network. Moreover, once the infrared target comes across long-term appearance changes in some sequences like “BusScale1” and “Exposure2”, the precision of other tackers is easy to decline dramatically. In contrast, our TSF-SiamMU tracker is adaptable to these appearance changes of the infrared target and performs well in these sequences owing to the multi-template update module. In conclusion, both quantitative and visualization results indicate that our TSF-SiamMU tracker outperforms other state-of-the-art methods not only on VOT-TIR 2016 dataset but also on GTOT dataset, which further demonstrates the credibility in various real infrared scenes of our proposed tracker.
 
\section{Conclusion}\label{section:5}
In this paper, we propose a twofold structured features-based Siamese multi-update tracker, called TSF-SiamMU. First of all, we design a novel feature fusion network to extract and make full use of both shallow spatial information and deep semantic information in a comprehensive manner, thereby providing richer and more robust feature representations for infrared target tracking. Further, a multi-template update module is proposed to effectively address the problem of tracking drift caused by the interference derived from infrared target appearance changes. Finally, both qualitative and quantitative experimental results on VOT-TIR 2016 and GTOT datasets demonstrate that our method achieves the balance of tracking accuracy and real-time tracking speed against other state-of-the-art trackers.

In the future work, we plan to give the Siamese tracker a broader sight in the search branch in order to deal with the more challenging tracking problems, for example, the wide range of target position changes in adjacent frames caused by camera motion.

\bibliographystyle{IEEEtran}
\bibliography{refs}

\vspace{-5mm}
\begin{IEEEbiography}
[{\includegraphics[width=1in,height=1.25in,clip,keepaspectratio]{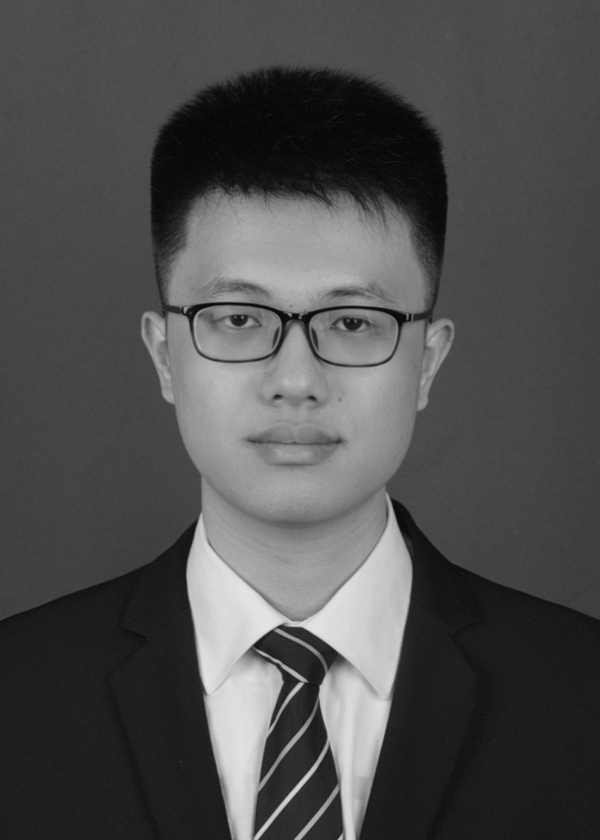}}] 
{Weijie Yan} received his B.S. degree in electronic science and technology from Nanjing University of Science \& Technology, Nanjing, China, in 2022. He is currently studying for the M.S. degree in physical electronics from Nanjing University of Science \& Technology, Nanjing, China. His main research interests include computer vision, machine learning and image processing.
\end{IEEEbiography}

\begin{IEEEbiography}
[{\includegraphics[width=1in,height=1.25in,clip,keepaspectratio]{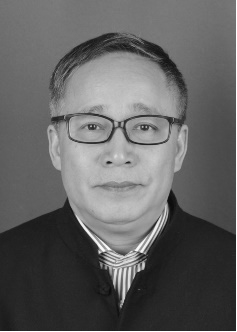}}] 
{Guohua Gu} received the B.S. and M.S. degrees in optical instrument from Nanjing University of Science \& Technology, Nanjing, China, in 1989 and 1996, respectively, and the Ph.D. degree in optical engineering from Nanjing University of Science \& Technology, Nanjing, China, in 2001. Since 2007, he has been a Professor with the School of Electronic and Optical Engineering, Nanjing University of Science \& Technology, Nanjing, China. His current research interests include optical design, computer version and machine learning.
\end{IEEEbiography}
\vspace{-5mm}

\begin{IEEEbiography}
[{\includegraphics[width=1in,height=1.25in,clip,keepaspectratio]{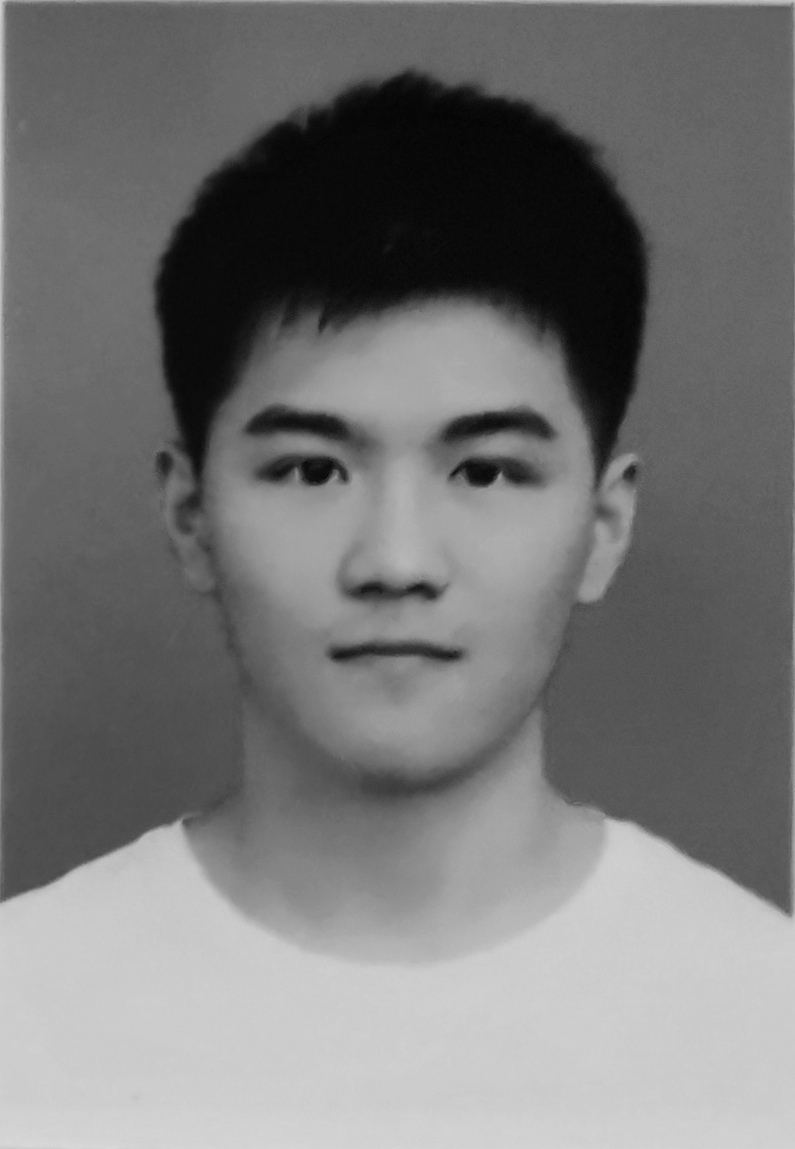}}] 
{Yunkai Xu} received his B.S. degree in electronic science and technology from Nanjing University of Science \& Technology, Nanjing, China, in 2020. He is currently studying for the Ph.D. degree in optical engineering from Nanjing University of Science \& Technology, Nanjing, China. His main research interests include computer vision and machine learning
\end{IEEEbiography}

\begin{IEEEbiography}
[{\includegraphics[width=1in,height=1.25in,clip,keepaspectratio]{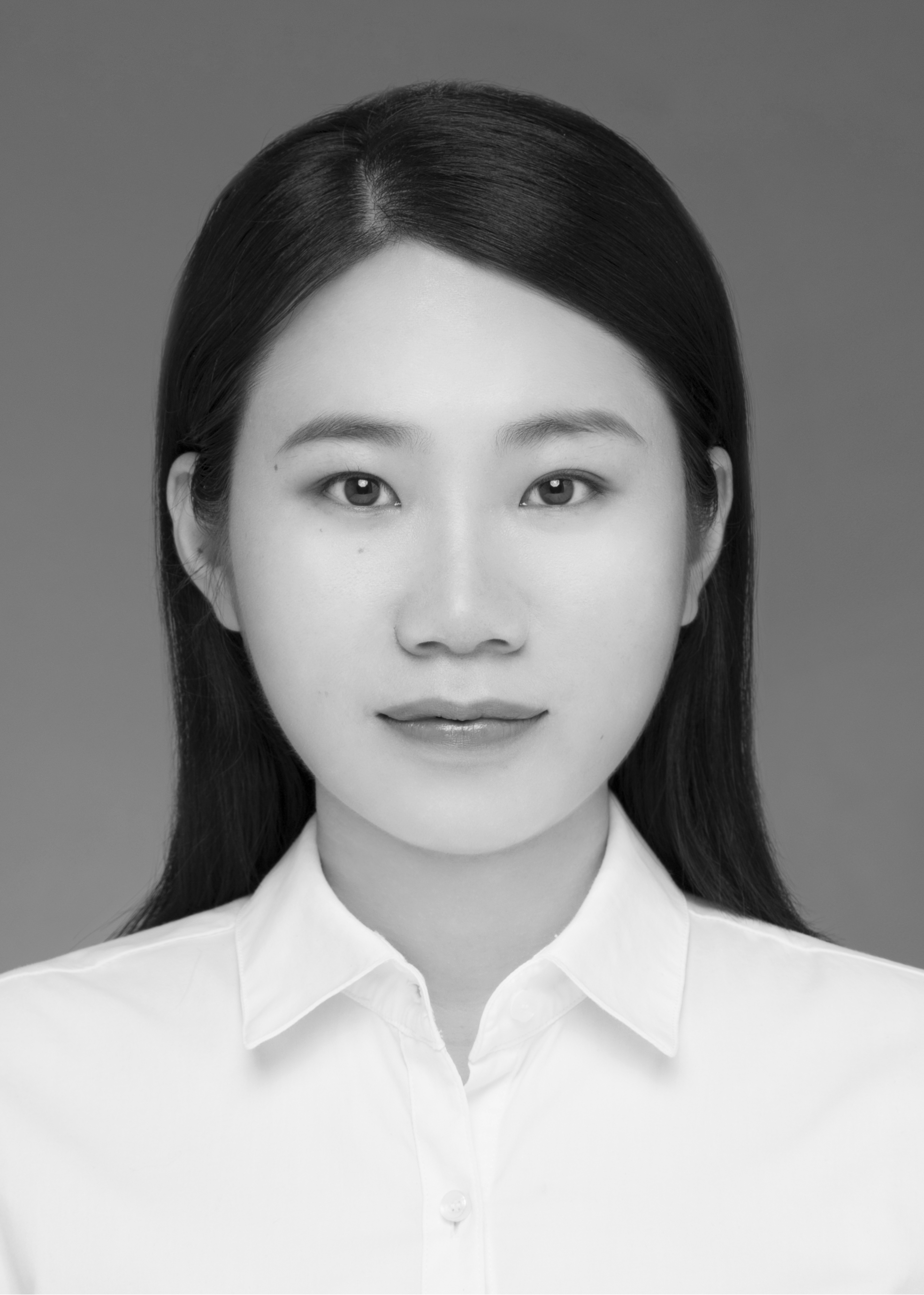}}] 
{Xiaofang Kong} received the B.S. degree in electronic information engineering and the Ph.D. degree in optical engineering from the Nanjing University of Science \& Technology, Nanjing, China, in 2013 and 2020, respectively.She is currently a Post-Doctoral Researcher with the National Key Laboratory of Transient Physics, Nanjing University of Science and Technology. Her research interests include dynamic parameter testing, and photoelectric detection and image processing.
\end{IEEEbiography}

\begin{IEEEbiography}
[{\includegraphics[width=1in,height=1.25in,clip,keepaspectratio]{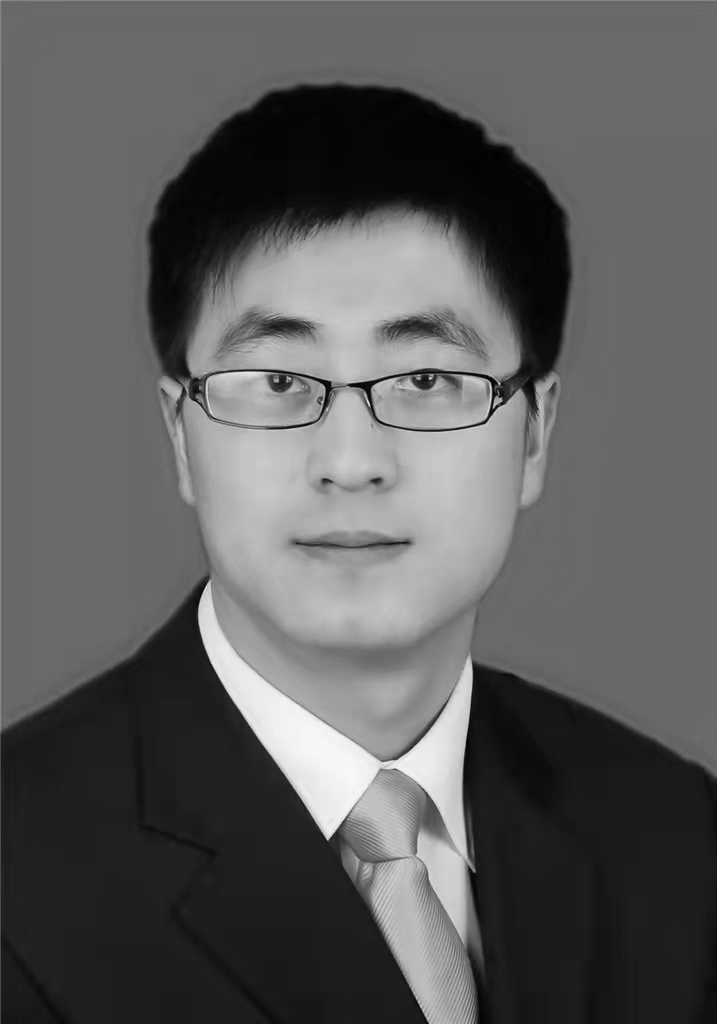}}] 
{Ajun Shao} received the B.S. degree in electronic science and technology and the M.S. degree in optical engineering from the Nanjing University of Science \& Technology, Nanjing, China, in 2012 and 2016, respectively, where he is currently pursuing
the Ph.D. degree in optical engineering. His main research interests include infrared imaging and image processing.
\end{IEEEbiography}
\vspace{5mm}

\begin{IEEEbiography}
[{\includegraphics[width=1in,height=1.25in,clip,keepaspectratio]{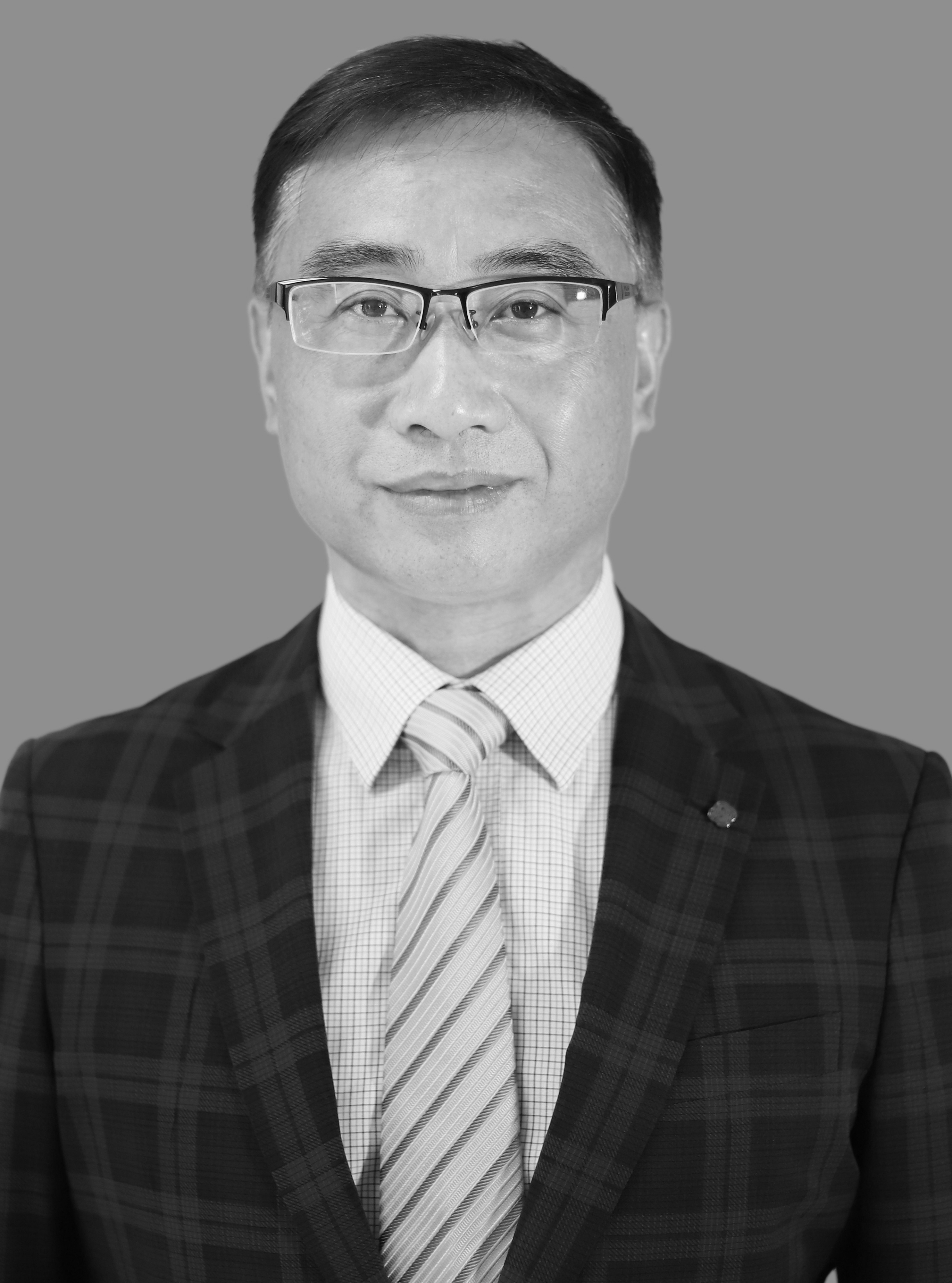}}] 
{Qian Chen} received the B.S. and M.S. degrees in optoelectronic technology from Nanjing University of Science \& Technology, in 1987 and 1991, respectively, and the Ph.D. degree in optical engineering from Nanjing University of Science \& Technology, Nanjing, China, in 1996. Since 1996, he has been a Professor with the School of Electronic and Optical Engineering, Nanjing University of Science \& Technology, Nanjing, China. His current research interests include optical design and computer version.
\end{IEEEbiography}
\vspace{5mm}

\begin{IEEEbiography}
[{\includegraphics[width=1in,height=1.25in,clip,keepaspectratio]{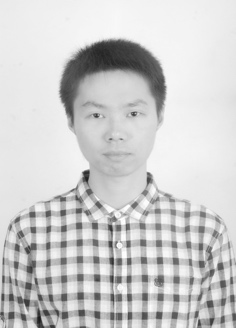}}] 
{Minjie Wan} received his B.S. degree in electronic science and technology from Nanjing University of Science \& Technology, Nanjing, China, in 2014 and the Ph.D. degree in optical engineering from Nanjing University of Science \& Technology, Nanjing, China, in 2020. He was a visiting Ph.D. student with the Department of Electrical and Computing Engineering, Université Laval, Quebec, Canada, from 2017-2018. He worked as a Post-Doctoral Researcher with the School of Electronic and Optical Engineering, Nanjing University of Science \& Technology, from 2020 to 2021, where he is currently an Associate Professor. His main research interests include image processing, computer vision, and computational imaging.
\end{IEEEbiography}

\vfill\pagebreak

\end{document}